\documentclass[fleqn,usenatbib]{mnras}

\usepackage{newtxtext,newtxmath}
\usepackage[T1]{fontenc}

\DeclareRobustCommand{\VAN}[3]{#2}
\let\VANthebibliography\thebibliography
\def\thebibliography{\DeclareRobustCommand{\VAN}[3]{##3}\VANthebibliography}

\usepackage{graphicx}	% Including figure files
\usepackage{amsmath}	% Advanced maths commands
\usepackage{microtype} % used to reduce spacing between text to fit in 5 pages

\usepackage{etoolbox, siunitx}
\robustify\bfseries

\usepackage{booktabs} % for tabular* environment

\usepackage[normalem]{ulem} % strikethrough text, e.g. \sout{insert text here}
\DeclareSIUnit\au{AU}
\DeclareSIUnit\Rsun{R_\odot}
\DeclareSIUnit\Rjup{R_\text{Jup}}
\DeclareSIUnit\Msun{M_\odot}
\DeclareSIUnit\Mjup{M_\text{Jup}}
\DeclareSIUnit\gyr{Gyr}
\DeclareSIUnit\ppt{ppt}
\DeclareSIUnit\ppm{ppm}

\newcommand*{\ra}[2][]{{% extra pair of braces to keep the \def-intion local!
    \def\SIUnitSymbolDegree{\textsuperscript{h}}%
    \def\SIUnitSymbolArcminute{\textsuperscript{m}}%
    \def\SIUnitSymbolArcsecond{\textsuperscript{s}}%
    \ang[#1]{#2}
    }%
}
\title[CM Draconis With TESS]{The Benchmark M Dwarf Eclipsing Binary CM Draconis With TESS: Spots, Flares and Ultra-Precise Parameters}

\author[Martin et al.]{%
        David V. Martin$^{1,2,3}$,
        Ritika Sethi$^{4,5}$,
        Tayt Armitage$^{2,6}$,
        Gregory J. Gilbert$^{7}$,\newauthor
        Romy Rodr\'iguez Mart\'inez$^{2,8}$,
        \& Emily A. Gilbert$^{9}$
\\
$^{1}$Department of Physics \& Astronomy, Tufts University, Medford, MA 02155, USA\\
$^{2}$Department of Astronomy, The Ohio State University, Columbus, OH 43210, USA\\
$^{3}$NASA Sagan Fellow\\
$^{4}$MIT Kavli Institute for Astrophysics and Space Research, Massachusetts Institute of Technology, Cambridge, MA 02139, USA\\
$^{5}$Department of Physical Sciences, Indian Institute of Science Education and Research, Berhampur, Odisha 760010, India\\
$^{6}$Department of Astronomy, University of Wisconsin Madison, WI 53706, USA\\
$^{7}$Department of Physics \& Astronomy, University of California Los Angeles, Los Angeles, CA 90095, USA\\
$^{8}$Centre of Astrophysics, Harvard University and Smithsonian, MA 02138, USA\\
$^{9}$Jet Propulsion Laboratory, California Institute of Technology, CA 91109, USA\\
*david.martin@tufts.edu\\
}

\date{First submitted to MNRAS Jan 25 2023, Re-submission Nov 29 2023, Accepted Dec 5 2023}

\pubyear{2024}

\begin{document}

\label{firstpage}
\maketitle

\begin{abstract}

A gold standard for the study of M dwarfs is the eclipsing binary CM Draconis. It is rare because it is bright ($J_{\rm mag}=8.5$) and contains twin fully convective stars on an almost perfectly edge-on orbit. Both masses and radii were previously measured to better than $1\%$ precision, amongst the best known. We use 15 sectors of data from the Transiting Exoplanet Survey Satellite (TESS) to show that CM Draconis is the gift that keeps on giving. Our paper has three main components. First, we present updated parameters, with radii and masses constrained to  previously unheard of precisions of $\approx 0.06\%$ and $\approx 0.12\%$, respectively. Second, we discover strong and variable spot modulation,  suggestive of spot clustering and an activity cycle on the order of  $\approx 4$ years. Third, we discover  163 flares. We find a relationship between the spot modulation and flare rate, with flares more likely to occur when the stars appear brighter. This may be due to a positive correlation between flares and the occurrence of bright spots (plages). The flare rate is surprisingly not reduced during eclipse, but one flare may show evidence of being occulted. We suggest the flares may be preferentially polar, which has positive implications for the habitability of planets orbiting M dwarfs.
\end{abstract}


%%%%%%%%%%%%%%%%%%%%%%%%%%%%%%%%%%%%%%%%%%%%%%%%%%

%%%%%%%%%%%%%%%%% BODY OF PAPER %%%%%%%%%%%%%%%%%%
%\section{SUMMARY}
%To date, planets around binary star systems are few and far between. The gravitational interactions between the two stars make for a hostile environment for planets after they form. Resonances in orbital periods between the binary and planet can create zones of instability that excite planet eccentricity and force ejection. We will use numerical simulations that include disks leftover from planet formation, planetary migration, and multi-planet systems (?) to test what kind of disk compositions, migration rates, and planet-planet interactions result in planets being parked in zones of stability, or elevated to zones of ejection. We can compare these results to observations to probe the accuracy and significance of the parameters included in our model, thus furthering our understanding of the positions and movements of circumbinary planets. 

\section{INTRODUCTION}\label{sec:introduction}

M dwarfs are the most common spectral type and very popular targets for exoplanet surveys, owing to a shorter period ``habitable zone'' that is easier to observe, and a high abundance of terrestrial planets \citep{dressing2015}. Significant time and resources have been dedicated to this field, including new spectrographs sensitive to redder wavelengths, transit surveys including Mearth, TRAPPIST, SPECULOOS and even TESS, as well as significant JWST time for transmission spectroscopy. However, there remain two major challenges with M dwarfs. 

First, the precision of our exoplanet parameters is a function of the precision in the host star parameters. For M dwarfs, empirical stellar mass-radius relations are poorly constrained \citep{torres2010} and often do not match theoretical models (e.g. radius ``inflation'', \citealt{Parsons2018}). This stems from a paucity of well-characterised M dwarfs in eclipsing binaries, which are the traditional calibrators \citep{Parsons2018,Gill2019,Swayne2021,Sebastian2023,Maxted2023}. 

A second challenge with M dwarfs is heightened stellar activity, evidenced by frequent flares and a high spot coverage. This impedes planet detection with transits \citep[e.g.][]{Kipping2017,Feliz2019,Gilbert2021,Gilbert2022} and radial velocities (RVs) \citep{Aigrain2016}. Activity also hinders characterisation of planets and their host star, and may be a cause of the discrepancy between models and observations of low-mass stars \citep{LopezMorales2005,Lubin2017}. Even if we were to discover an ``Earth-like'' planet, its habitability may be questionable if the planet is frequently bombarded by flares and associated coronal mass ejections (CMEs) \citep{Ranjan2017,Tilley2019,France2020,Bogner2022}. The largest well-recorded solar storm (flare + CME) for the Sun was the Carrington Event of 1859. It released $\approx10^{32}$ erg, causing aurorae to be seen globally. If it occurred today it would devastate our power grids. Carrington-esque events can be a daily occurrence on M dwarfs. More knowledge of M-dwarf activity is needed to assess the viability of potentially habitable worlds. Fortunately, the availability of long-term photometry from observatories such as Kepler, TESS and the All Sky Automated Survey for SuperNovae (ASAS-SN) has revolutionised studies of flares and spots \citep{Davenport2016,Schmidt2019,Gunther2020,Rodriguez2020,feinstein2022,Mendoza2022}. 

In this paper we use  15 sectors of TESS photometry to study CM Draconis, a  double-lined, spectroscopic eclipsing binary with twin M dwarfs, introduced in detail in Sect.~\ref{sec:cmdra}. There are two fundamental aspects to our paper, related to the aforementioned two challenges with M dwarfs. First, we use the TESS photometry and archival radial velocities (Sect.~\ref{sec:data_acquisition}) to provide an order of magnitude improved fit to the eclipses (Sect.~\ref{sec:eclipse_fitting}), which we compare to stellar models. Second, we identify and characterise CM Dra's activity through  flares (Sect.~\ref{sec:flares}) and star spots (Sect.~\ref{sec:spots}). We quantify the flare rate and  place constraints on both the longitudinal and latitudinal distribution of the stellar activity. Through this, we quantify the connection between spots and flares and estimate the activity cycle, akin to the 11-year Solar cycle. All of our analysis is contained in Sect.~\ref{sec:analysis}, before concluding in Sect.~\ref{sec:conclusion}.

%To better understand M dwarf activity, and indeed its impact on habitable planets, we need to know the latitudinal distribution of flares and CMEs. On the Sun flares, CMEs \& spots are typically equatorial ($\pm 30^{\circ}$), following an 11 year cycle \citep{Zhang2007}. For M dwarfs there is preliminary evidence that spots \citep{Barnes2017} and flares \citep{Ilin2021} might be polar. This would likely be good for habitability, since the flux from the flares would be attenuated and the charged particles from flares would be directed away from the planet. Note that this assumes spin-orbit alignment for the star and planet, which works for the solar system and most exoplanets.

\section{THE CM DRACONIS SYSTEM}\label{sec:cmdra}

\begin{figure*}
    \centering
    \includegraphics[width=0.99\textwidth]{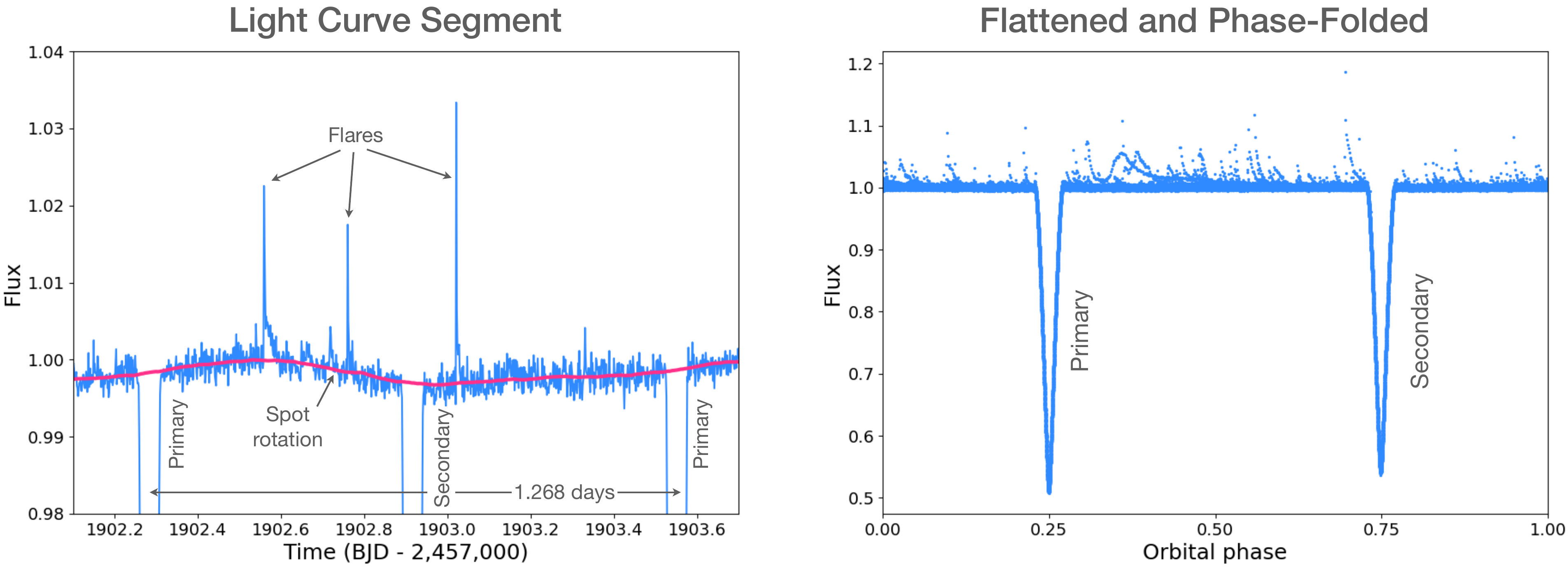}
    \caption{{\bf Left:} segment of PDCSAP photometry, illustrating the four major components of the CM Dra light curve: primary eclipses, secondary eclipses, flares and out-of-eclipse modulation, the last of which is predominantly due to spots on the rotating stars. {\bf Right:} CM Dra light curve that has been detrended (i.e. flattened) and phase-folded on the orbital period of 1.268 days. The primary and secondary depths are similar, almost corresponding to a 50\% reduction in flux because the orbital alignment is almost perfectly edge-on, and the two stars are almost identical. Flares are seen to occur at all orbital phases (analysed in depth in Sect.~\ref{subsubsec:orbital_phase}).}
    \label{fig:light curve}
\end{figure*}

CM Draconis (CM Dra henceforth) is a benchmark system for the study of M dwarfs. It is a 1.26 day eclipsing binary of near twin fully convective dM4.5 stars. The two stars are remarkably well aligned to our line of sight with ($I=89.8^{\circ}$, $b = 0.12$). With near identical radii ($R_{\rm A} = 0.251$ and $R_{\rm B} = 0.238$) this means that both primary and eclipses result in a near perfect occultation (50\% drop in total flux). With near equal masses ($M_{\rm A}=0.225$ and $M_{\rm B} = 0.211$) CM Dra is a double-lined spectroscopic binary, meaning that model-independent masses can be derived for both components. Indeed, the seminal \citet{torres2010} stellar mass-radius relation, constructed using the best-characterised stars, contains only two fully convective stars: CM Dra A and CM Dra B. This is problematic not only due to a small sample size, but because both stars are ``inflated'' by $\approx 5\%$, meaning that their radius is higher than expected from theoretical models \citep{Feiden2014inflation}. The stars therefore may not be representative of typical low-mass stars. We will re-assess the inflation of CM Dra in Sect.~\ref{subsec:inflation} (spoiler alert: they are still inflated).

CM Dra is actually a triple star system, with a bound white dwarf WD 1633+572 located 25.7'' away, corresponding to a separation of at least 370 AU \citep{Deeg2008}. The white dwarf, by virtue of the known relationship between age and effective temperature, provides an age estimate of the system of $8.5\pm3.5$ Gyr \citep{Feiden2014age}. CM Dra also has sub-solar metallicity: [Fe/H]$=-0.30\pm0.12$ from \citet{Terrien2012}. We list observable, physical and orbital parameters in  Table~\ref{tab:summary}, including both our new values (Sect.~\ref{sec:eclipse_fitting}) and literature values from the previous reference \citep{morales2009}\footnote{We use the `Average' column in their Table 5.}.

\begin{table}
%    \renewcommand{\arraystretch}{1.2}
%    \sisetup{round-mode=places}

    \centering
    \caption{Observational properties of the CM Draconis double M dwarf eclipsing binary. Data taken from the TESS Input Catalog v8.2. TESS sectors refer to current and predicted availability as of November 2023}  %title of the table
%    \resizebox{\columnwidth}{!}{
    % align at dash later
    % https://tex.stackexchange.com/questions/4964/align-equals-sign-in-table
    \footnotesize
    \begin{tabular*}{\columnwidth}{@{\extracolsep{\fill}}
        % l
        l
        l
        l
        }
        \toprule
        \toprule
        Parameter & Description & Value \\
        \midrule

        \multicolumn{3}{@{}l@{}}{\emph{Observable Properties}} \\[5pt]
        ${\rm TIC}$ & TESS Input Catalog & 199574208 \\
        ${\rm GAIA}$ & ID & 1431176943768690816 \\
        ${\rm 2MASS}$ & ID & 16342040+5709439 \\
        $\alpha$ & Right ascension & $248.58470795746^{\circ}$  \\
        & & (\ra[angle-symbol-over-decimal,minimum-integer-digits=2]{16;34;20.330}) \\
        $\delta$ & Declination & $+57.1623237628578^{\circ}$ \\
        & & (\ang[angle-symbol-over-decimal]{+57;09;44.37}) \\
        TESS sectors & Cycle 2 & 16, 19, 22, 23, 24, 25, 26 \\
        & Cycle 4 &  49, 50, 51, 52 \\
        & Cycle 5 &   56, 57, 58, 59 \\
        & Cycle 6 (future$^*$) &   76$^*$,77$^*$,78$^*$,79$^*$,80$^*$,82$^*$,83$^*$ \\
        $V_\mathrm{mag}$ & Apparent V magnitude & 13.35 \\
        $J_\mathrm{mag}$ & Apparent J magnitude & 8.501 \\
        $d$  & Distance (pc) & $14.8436\pm0.01145$ \\
        
                \midrule
        \multicolumn{3}{@{}l@{}}{\emph{Primary Star Properties - CM Dra A}} \\[5pt]
        %$M_{\rm A}$  & Mass ($M_{\odot}$) & $0.22501 \pm 0.00025$ \\
                $M_{\rm A}$  & Mass ($M_{\odot}$) & $0.22507 \pm 0.00024$ \\
          & \citet{morales2009} & $0.2310\pm0.0009$ \\
        %$R_{\rm A}$  & Radius ($R_{\odot}$) & $0.25104 \pm 0.00017$  \\
        $R_{\rm A}$  &Radius ($R_{\odot}$) & $0.25113 \pm 0.00016$  \\
          & \citet{morales2009} & $0.2534\pm0.0019$ \\
        %$\log g_{\rm A}$  & Surface gravity (cgs) & $4.994\pm0.007$ \\
        
        \midrule
        \multicolumn{3}{@{}l@{}}{\emph{Secondary Star Properties - CM Dra B}} \\[5pt]
        %$M_{\rm B}$  & Mass ($M_{\odot}$) & $0.21011 \pm 0.00027$ \\
        $M_{\rm B}$  & Mass ($M_{\odot}$) & $0.21017 \pm 0.00028$ \\
          & \citet{morales2009} & $0.2141\pm0.0010$ \\
        %$R_{\rm B}$  & Radius ($R_{\odot}$) & $0.23745 \pm 0.00015$ \\
        $R_{\rm B}$  &Radius ($R_{\odot}$) & $0.23732 \pm 0.00014$ \\
          & \citet{morales2009} & $0.2396\pm0.0015$ \\
        %$\log g_{\rm B}$  & Surface gravity (cgs) & $5.009\pm0.006$ \\
        s & Surface brightness ratio & $0.98133
\pm 0.00081$ \\
        
        \midrule
        \multicolumn{3}{@{}l@{}}{\emph{Binary Orbital Properties}} \\[5pt]
        $a$ & Semi-major axis  (AU)  & $0.0173945\pm 0.0000064$ \\
         & \citet{morales2009}  & $0.01752\pm0.00021$ \\
        $P$ & Period (day) & $1.2683900573\pm 0.0000000017$ \\
         & \citet{morales2009} & $1.268389985\pm0.000000005$ \\
        $e$ & Eccentricity & $0.00527\pm0.00021$ \\
        & \citet{morales2009} & $0.0054\pm0.0013$ \\
        $\omega$ & Argument of periapse ($^{\circ}$) & $107.98\pm0.70$ \\
         & \citet{morales2009}  & $107.6\pm6.3$ \\
        $I$ & Sky inclination ($^{\circ}$) & $89.5514
\pm0.0020$ \\
 & \citet{morales2009} & $89.769
\pm0.073$ \\
        $b$ & Impact parameter & $0.11711
\pm0.00054$ \\
 & \citet{morales2009} & $0.060\pm0.018$ \\
        
        \bottomrule
    \end{tabular*}
    \label{tab:summary}
\end{table}

CM Dra was the first target of circumbinary planet surveys \citep{schneider1994,jenkins1996,deeg1998,deeg2000}, with searches based on both eclipse timing variations and planetary transits. Unfortuantely, no planets have been found orbiting CM Dra, but over a dozen circumbinary planets have been since discovered (review in \citealt{martin2018}). CM Dra being planetless is consistent with the \citet{martin2014,armstrong2014} discovery that the tightest eclipsing binaries ($P_{\rm bin} \lesssim 7$ days) lack planets. This was interpreted by \citet{munoz2015,martin2015,hamers2016} as evidence for a ``violent'', high-eccentricity evolution history of such tight binaries, which would destabilise planets. However, whilst circumbinary planets are objectively cool \citep{martin2021}, they are not the focus of this paper.

It has also been known for decades that CM Dra exhibits flares.  \citealt {Lacy1977} conducted a dedicated photometric flare search on CM Dra with 18 hours of data spread over one year, covering all orbital phases with a high-speed cadence. No flares were observed in this part of the survey. However, there was also a shorter survey of differential near-infrared photometry with a different goal of improving the limb darkening and radius measurements. To quote \citet{Lacy1977}, ``Due to the perversity of Nature, the only flare we have detected occurred during the infrared differential photometry when it was least expected and least desired!'' Based on this single event, the flare rate was estimated to be between 0.48 and 1.2 flares per day. Subsequent studies by \citet{Metcalfe1996,kim1997,Nelson2007} calculated similar flare rates based on samples of typically less than ten. \citet{Stelzer2022} discovered 16 flares in one TESS sector. In our paper we will demonstrate the detection of  163 flares in 15 TESS sectors.

\section{DATA ACQUISITION AND PREPARATION}\label{sec:data_acquisition}

%\subsection{TESS PHOTOMETRY}\label{sec:tess_photometry}

In Fig.~\ref{fig:light curve} (left) we show a cutaway of the light curve which highlights four phenomena: 1: primary eclipses; 2: secondary eclipses; 3: out of eclipse periodic variability (which we attribute to spots) and 4: flares. In Fig.~\ref{fig:light curve} (right) we show the light curve phase-folded on the orbital period of 1.28 days. In the latter plot, the out of eclipse variability has been flattened but the flares and eclipses remain.

For radial velocities, we do not take new measurements, but we instead use \citet{morales2009}'s RV measurements, which come from an improved analysis of the \citet{Metcalfe1996} spectroscopy. There are 233 measurements for each star, with a median precision of 1.2 km/s and 1.4 km/s for the primary and secondary star, respectively. With updated spectrographs and improved binary spectroscopy (e.g. \citealt{Standing2022}) we could improve upon these RV's, but we leave that to a future study.

\section{COMBINED PHOTOMETRY \& RV FIT}\label{sec:eclipse_fitting}

%\begin{figure*}
%    \centering
%    \includegraphics[width=0.99\textwidth]{orbital_phase.pdf}
%    \caption{}
%    \label{fig:light curve}
%\end{figure*}

\begin{figure}
    \centering
    \includegraphics[width=0.50\textwidth]{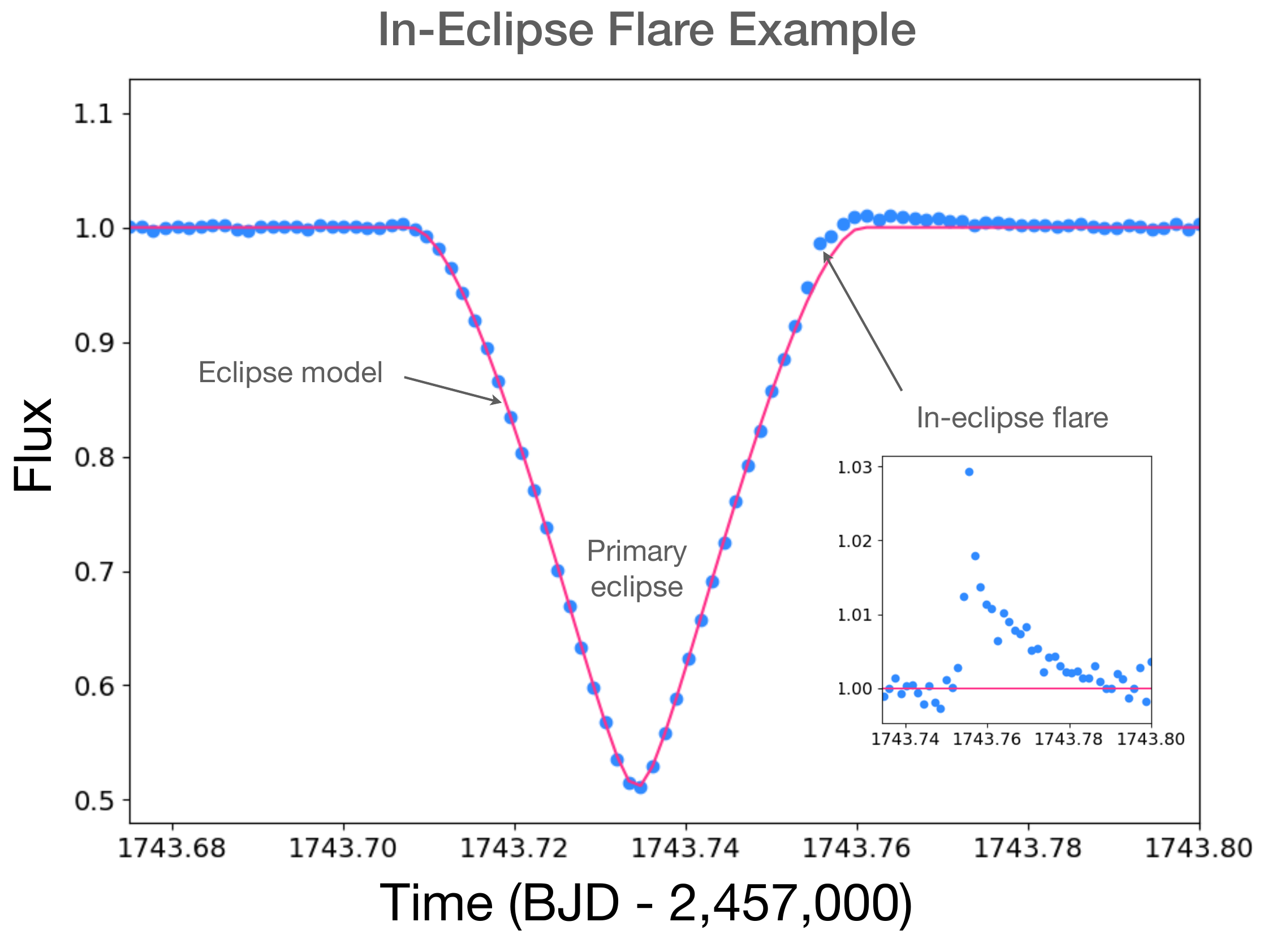}
    \caption{Eclipse model fitted with \textsc{exoplanet} (pink line) to the flattened TESS data (blue data). Since there are well over 100 eclipses, the occasional flare that coincides with an eclipse does not affect the model. Therefore, when the eclipse model is subtracted the flare stands out (inset).}
    \label{fig:in_eclipse_flare_example}
\end{figure}

\begin{figure*}
    \centering
    \includegraphics[width=0.99\textwidth]{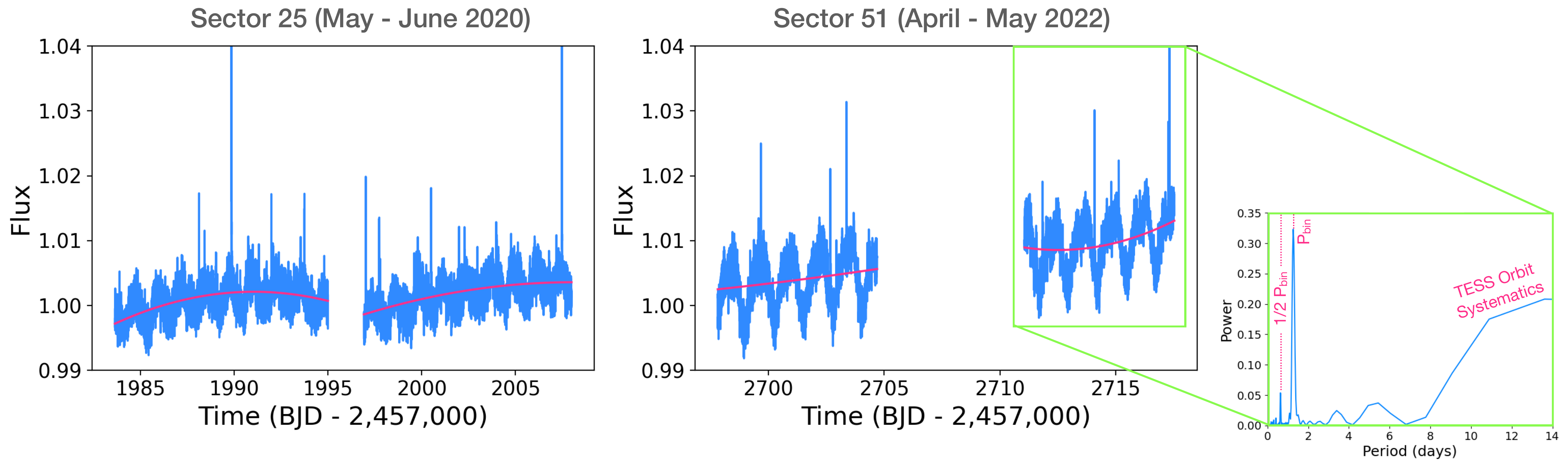}
    \caption{TESS PDCSAP photometry of CM Dra where the only processing is the eclipses have been cut out. On the {\bf left} is sector 25 and on the {\bf right} is sector 51, spaced roughly two years apart. Each sector contains two 13.7-day orbits of the TESS spacecraft. The central gap is when the data is downloaded to Earth. Within each TESS orbit is a parabolic trend, which we fit and remove (pink line). The remaining trends we attribute to spot modulation and possibly ellipsoidal variation. The inset shows a Lomb Scargle periodogram for the second TESS orbit in sector 51. The strongest periodicity is at $P_{\rm bin}$, followed by the longer period parabolic trend and a smaller peak at $1/2P_{\rm bin}$.}
    \label{fig:parabolas}
\end{figure*}

There are several steps to preparing the light curve. We provide online the light curve at every stage of the processing. 

We first detrended and stitched the TESS data using the \textsc{Wotan} package \citep{hippke2019b}. We use the default Tukey's bi-weight filter, which \citet{hippke2019b} has shown to be the best at preserving the shape of short-timescale events (like eclipses and flares) whilst removing longer trends (e.g. spot modulation). We use a window length of 0.228 days, which is four times the eclipse width. Two exceptions to this were for massive flare events at times 1953 and 2726 (BJD - 2,457,000), for which this \textsc{Wotan} filter affects the flare shape. We account for this by taking the \textsc{Wotan} trend from the previous 1.268 days (the orbital period) and using that to detrend the light curve over this flare. The end result of this process is a flattened light curve with only eclipses and flares. 

The next step was to fit the flattened photometry using the \textsc{exoplanet} software \citep{foreman-mackey2021}. This incorporates an eclipsing binary model from \textsc{starry} \citep{luger2019} with limb darkening parameters from \citet{kipping2013}.  The purpose of this initial fit was not to derive stellar and orbital parameters, but rather to create a light curve model that we could subtract to remove the eclipses. Doing so produces a flat light curve with nothing but flares. By model-subtraction of the eclipses rather than simply cutting them (as was the case in \citealt{Stelzer2022}), we can illuminate in-eclipse flares, as demonstrated in Fig.~\ref{fig:in_eclipse_flare_example}. This means we can compare the flare rate in and outside of eclipse (Sect.~\ref{subsubsec:orbital_phase}) and search for occulted flares which would help us to constrain the location of the flaring region on the star (Sect.~\ref{subsec:occulted_flares}). Subtraction of the eclipses also maintains $\sim 10\%$ of the data that would otherwise be masked out.

We then used the \textsc{stella} neural network software \citep{feinstein2020stella} to identify flares in this flattened, eclipse-subtracted light curve (detailed in Sect.~\ref{sec:flares}). We use this list of flare times and durations to cut them from the flattened light curve. We are finally left with a light curve that is flattened and contains nothing but eclipses and a few gaps where there used to be flares (both in and outside of eclipse). 

On this light curve we create a joint photometry \& RV fit with \textsc{exoplanet}. The free parameters in the model are the binary orbital period $P_{\rm bin}$,, total baseline system flux $F_0$, the mass $M_A$ and radius $R_A$ of the primary star, primary-to-secondary mass ratio $q$, radius ratio $r$, and surface brightness ratio $s$, epoch of first eclipse $t_0$, impact parameter $b$, and eccentricity vectors $(\sqrt{e}\sin\omega, \sqrt{e}\sin\omega)$. All prior distributions were chosen to be uninformative (i.e. uniform on $e$ and $b$; normal on $F_0$, $P_{\rm bin}$, $t_0$, $q$, $r$, and $s$; log-normal on $M_A$ and $R_A$). Each star's quadratic limb darkening profile was parameterized using the uninformative prescription of \citet{Kipping2013b}.

After first estimating the maximum a posteriori parameters, we use \textsc{PyMC3} to derive a posterior distribution and $1\sigma$ error bars, provided in Table~\ref{tab:summary}.

Our radius measurement precision is 0.064\% and 0.059\% for the primary and secondary star, respectively.  These measurements represent an order of magnitude improvement from the previous \citet{morales2009} precisions of 0.75\% and 0.63\%, which were already some of the best known for M dwarfs. This is a product of the exquisite precision and long baseline of the TESS photometry.

Our precision on the stellar masses is 0.11\% and 0.12\% for A and B, respectively. This is an improvement with respect to the \citet{morales2009} precisions of 0.39\% and 0.47\%. The improvement is not as large as it was with the radius, but that is because we improved the photometry but used the same RVs. The fact that the masses have improved despite using the same RVs is due to improved orbital parameters. For example, the precise eclipse phase from photometry more precisely constrain the eccentricity than the RVs alone. Since the conversion from RV semi-amplitude $K$ to mass has an eccentricity dependence, an improved measurement of $e$ improves the measurement of $M_{\rm A}$ and $M_{\rm B}$.

Overall, these are the most precise M dwarf parameters ever measured and are likely amongst the most precisely-measured stars ever. There is some discrepancy between our measurements and the \citet{morales2009} measurements. Based on the \citet{morales2009} measurement error, our radii are both $\approx 1.3\sigma$ smaller. Our masses are $6.6\sigma$ and $4\sigma$ smaller for A and B, respectively. Compared with \citet{morales2009}, we use new photometry but the same radial velocities. It may therefore seem surprising that we get different masses. However, as previously mentioned, changes to the photometry will still impact the masses. We have moved from ground-based photometry to an order of magnitude more space-based data.   The source of the mass discrepancy might be a discrepancy in the derived impact parameter, since the radial velocity minimum mass becomes a true mass when the inclination (i.e. impact parameter) is measured. Accurate, unbiased and precise determination of the impact parameter is known to be challenging, in particular for grazing transits/eclipses, as is the case for CM Dra given the stars are practically twins \citep{Gilbert2022,Gilbert2022b}. This can be particularly the case when comparing derived values using different instruments, since one also has to factor in different prescriptions for limb darkening. 

Another possible explanation is that the stars have physically changed over the decades since the observations used in  \citet{morales2009}. We know that CM Dra is active, with spots evolving on what we will derive to be a $\approx 4$-year timescale (Sect.~\ref{subsec:activity_cycle}). Subtle changes in the spot distribution inside and outside eclipse will change the derived physical parameters . A more complex physical model that includes both the eclipsing bodies and a modelled spot distribution, e.g. using \textsc{PHOENIX} \citep{Conroy2020}, is being the scope of our paper.

Finally, as we will discuss in Sect.~\ref{subsec:inflation}, the degree to which the CM Dra stars have inflated radii with respect to models has not changed with our fit.

\section{FLARE DETECTION}\label{sec:flares}

\begin{figure*}
    \centering
    \includegraphics[width=0.99\textwidth]{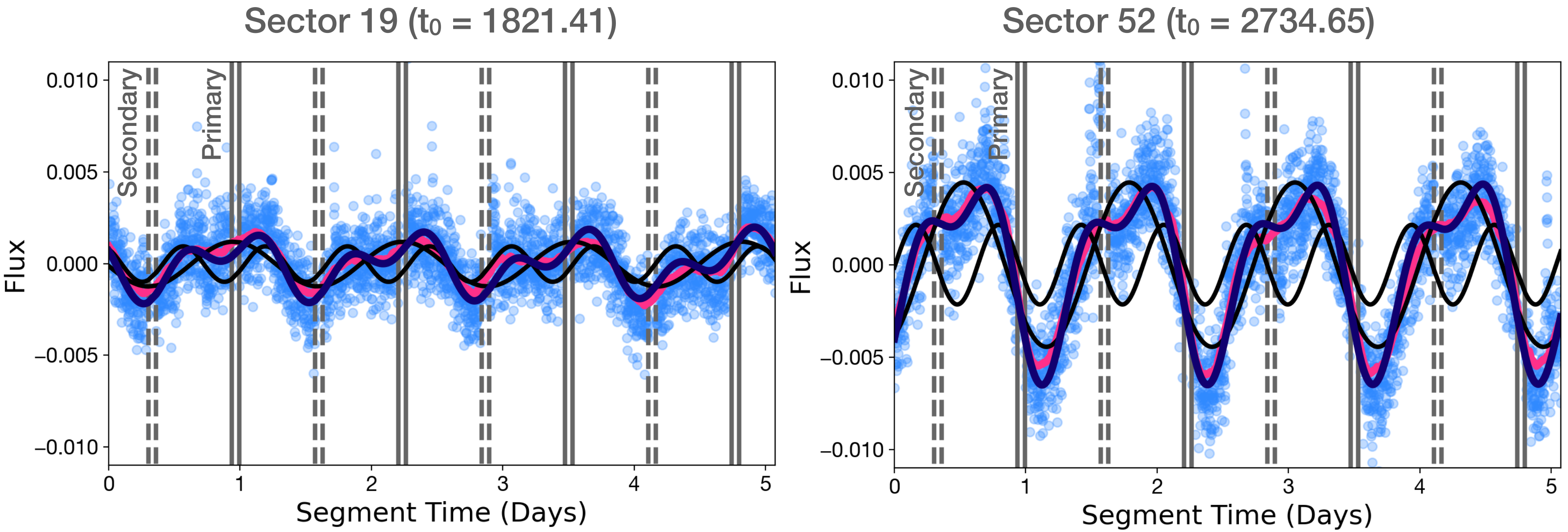}
    \caption{ Two example lightcurve segments: one from TESS sector 19 (left) and one from TESS sector 52 (right), over 900 days later. Each segment has a length equal to four orbital periods of CM Dra ($4\times1.268=5.072$ days). Both types of eclipses have been cut out, with solid and dashed vertical lines used to denote when the primary and secondary eclipses occur, respectively. The pink curves indicate the lightcurve trend calculated using a Tukey's biweight filter in \textsc{Wotan}. Within each segment we create a fit to the data with a pair of sinusoids: one with a period near $P_{\rm orb}$ and one with a period half that. This fitted curve is shown in dark blue. In both segments it is similar to the pink \textsc{Wotan} trend, indicating that our two-sinuisoid model is appropriate and well-fitted. By comparing the two sectors, it is clear that the amplitude and phase of both sinuisoids has changed significantly, indicating substantial evolution in the spots over this $\sim900$ day timespan.}
    \label{fig:fitted_sine_curves}
\end{figure*}

We detect flares on the flattened, eclipse-subtracted light curve using the \textsc{stella} software \citep{feinstein2020stella}. This is a flare detection algorithm based on a Convoluted Neural Network (CNN), and has been applied to TESS  detection in \citet{feinstein2020b,feinstein2022}. We use the default training set. \textsc{stella} assigns a ``flare likelihood probability'' to each flare, for which the default threshold is 50\%. These probabilities are provided in our online table of flares. 

We detect 175 flares above this default threshold. However, within this paper we instead use a more conservative threshold of 75\%, from which {\it we detect   163 flares}. There are several reasons for a stricter threshold. First, by eye many of the lower probability flares look unconvincing above typical noise. Second, when calculating the flare rate in Sect.~\ref{fig:flare_frequency_distribution} we are calculating the flare frequency distribution, with accounts for incomplete flare detection at low energies, i.e. the flares likely to have low probability from \textsc{stella}. Finally, when calculating the distribution of flares (Sect.~\ref{subsec:flare_phases}) we prefer a conservative approach where the flare statistics may be weaker but we are more confident that they are actually flares.

We follow \citet{Shibayama2013,Gunther2020} to calculate the energy of each flare. The flare is modelled as black body radiation with $T=9000$ K. Given we do not know which star a given flare is occurring on, and they are close to identical, we take the stellar radius and temperature to be the average between stars A and B. In general, the luminosity of a blackbody with in a given observing bandpass (i.e. TESS) is calculated by

\begin{equation}
    L' = A\int R_\lambda B_\lambda(T)d\lambda,
\end{equation}
where $A$ is the surface area of the blackbody, $R_\lambda$ is the TESS response function, combining both the detector filter transmission and the quantum efficiency, and $B(T)$ is the Planck function as a function of the temperature of the blackbody temperature $T$. This equation, evaluated for the flare and star, is

\begin{align}
    L_\star' &= \pi R_\star^2 \int R_\lambda B_\lambda(T_{\rm eff})d\lambda, \\
    L_{\rm flare}'(t) &= A_{\rm flare}(t) \int R_\lambda B_\lambda(T_{\rm flare})d\lambda,
\end{align}
where the above equation demonstrates that the flare luminosity and hence area will be time-dependent, compared with the constant, quiescent stellar luminosity.

In the normalised light curve the change in flux is related to the change in luminosity by

\begin{equation}
   \frac{L_{\rm flare}'(t)}{L_\star'} =  2\left(\frac{\Delta F}{F}\right)(t)
\end{equation}
The factor of 2 differs from \citet{Shibayama2013,Gunther2020} to account for the fact that CM Dra contains two stars of nearly equal brightness. For example, a 10\% increase in the light curve flux corresponds to a 20\% increase in the brightness of the individual flaring star. 

We  therefore solve for the area of the flare region:
\begin{align}
    A_{\rm flare}(t) &= \left(\frac{\Delta F}{F}\right)(t)\pi R_{\rm star}^2 \frac{\int R_\lambda B_\lambda(T_{\rm eff})d\lambda}{\int R_\lambda B_\lambda(T_{\rm flare})d\lambda} \\
    A_{\rm flare}(t) &= 0.0203\left(\frac{\Delta F}{F}\right)(t)\pi R_{\rm star}^2
\end{align}

The bolometric flare luminosity is given by
\begin{equation}
    L_{\rm flare}(t) = \sigma_{\rm SB}T_{\rm flare}^4A_{\rm flare}(t),
\end{equation}
where $\sigma_{\rm SB}$ is the Stefan-Boltzmann constant. By integrating the luminosity over the flare time we obtain the total flare energy:

\begin{equation}
    E_{\rm flare} = \int L_{\rm flare}(t)dt.
\end{equation}

Our 163 detected flares span energies between $9.1\times10^{30}$ and $2.4\times 10^{33}$ erg, as discussed in Sect.~\ref{fig:flare_frequency_distribution}. The largest flare area is $\approx 7\times10^{12} m^2$, which corresponds to a diameter that is $\approx 10\%$ that of the star. This is a much larger spot-to-star ratio than we see on the Sun, but smaller than the highly active M-dwarfs studied by \citet{Ilin2021}.

\section{Spot detection}\label{sec:spots}

CM Draconis is a so-called ``BY Draconis'' variable, which is a main sequence variable (typically a K or M dwarf) exhibiting variations on the order of roughly a magnitude. The variation is caused by the presence of star spots. Since spots evolve and may exist at different latitudes (and hence different rotation rates), this class of variable is only ``semi-regular''. This is in contrast with say cepheid and ellipsoidal variables, which are ``fully regular''.

Since CM Dra is such a tight binary, we expect the two stars to be 1:1 spin-orbit synchronised due to tides. We also expect the effect of spots to be visible on CM Dra given M dwarfs are highly active and it is a bright target. Overall, we expect an out of eclipse variability with a period of 1.268 days. 

In Fig~\ref{fig:parabolas} we plot sectors 25 and 51 of TESS. They are separated by approximately two years. Each sector contains two of TESS's 13.7-day orbits around the Earth. There is a gap in the middle of each sector for data download\footnote{This gap is larger in sector 51 because of a change to the TESS mission to take more short cadence data, which takes longer to download to Earth.}. For clarity the eclipses have been removed. Several things are visibly apparent. First, there is a modulation of the light curve with a periodicity a bit longer than 1 day, as expected. However, it is more complicated than a simple sinusoid. Within each TESS orbit there is a longer term parabolic modulation of the light curve (fitted pink line). This changes from TESS orbit to orbit. Given this occurs on a TESS orbital timescale, we attribute it to systematics in the data and not something physical with CM Dra. 

Even if we account for these TESS systematics, it is visible that the shape of the photometric modulation is not strictly a single sinusoid. This is emphasised in the Lomb-Scargle periodogram (Fig.~\ref{fig:parabolas} inset) for the second TESS orbit in sector 51, where we see the three most significant powers are i) $P_{\rm bin}$, ii) $\approx 14$ days due to the parabolic trend of TESS systematics, and iii) $1/2P_{\rm bin}$. Finally, the amplitude of the modulation is visibly larger in sector 51 than it is in 25, suggesting that the spot coverage has become more pronounced and/or uneven over time.

To quantify the out-of-eclipse light curve modulation, and its time evolution, we invoke the following procedure:
\begin{enumerate}
    \item Cut (not subtract) all eclipses from the light curve.
    \item Split the data up into separate TESS orbits and fit and subtract a parabola (like Fig.~\ref{fig:parabolas}). In some TESS orbits there is a sharp change to the flux near the start or end of the orbit, beyond the simple parabolic trend, in which case we manually cut these edge data.
    \item Further split the data up into $4\times 1.268$ day segments, where 1.268 days is CM Dra's orbital period. The choice of spitting the data up into $4\times P_{\rm bin}$ was an ad hoc compromise between having a sufficiently long timeseries to make a good sinusoid fit and not extending the timeseries too long such that we would blur what we will see is fairly rapid spot evolution.
    \item Within each segment, we fit two sinusoids using \textsc{scipy.optimize.curve\_fit}, using $1.268$ and $0.634$ days as the initial period guesses (like Fig.~\ref{fig:fitted_sine_curves}).
\end{enumerate}

\begin{figure}
    \centering
    \includegraphics[width=0.50\textwidth]{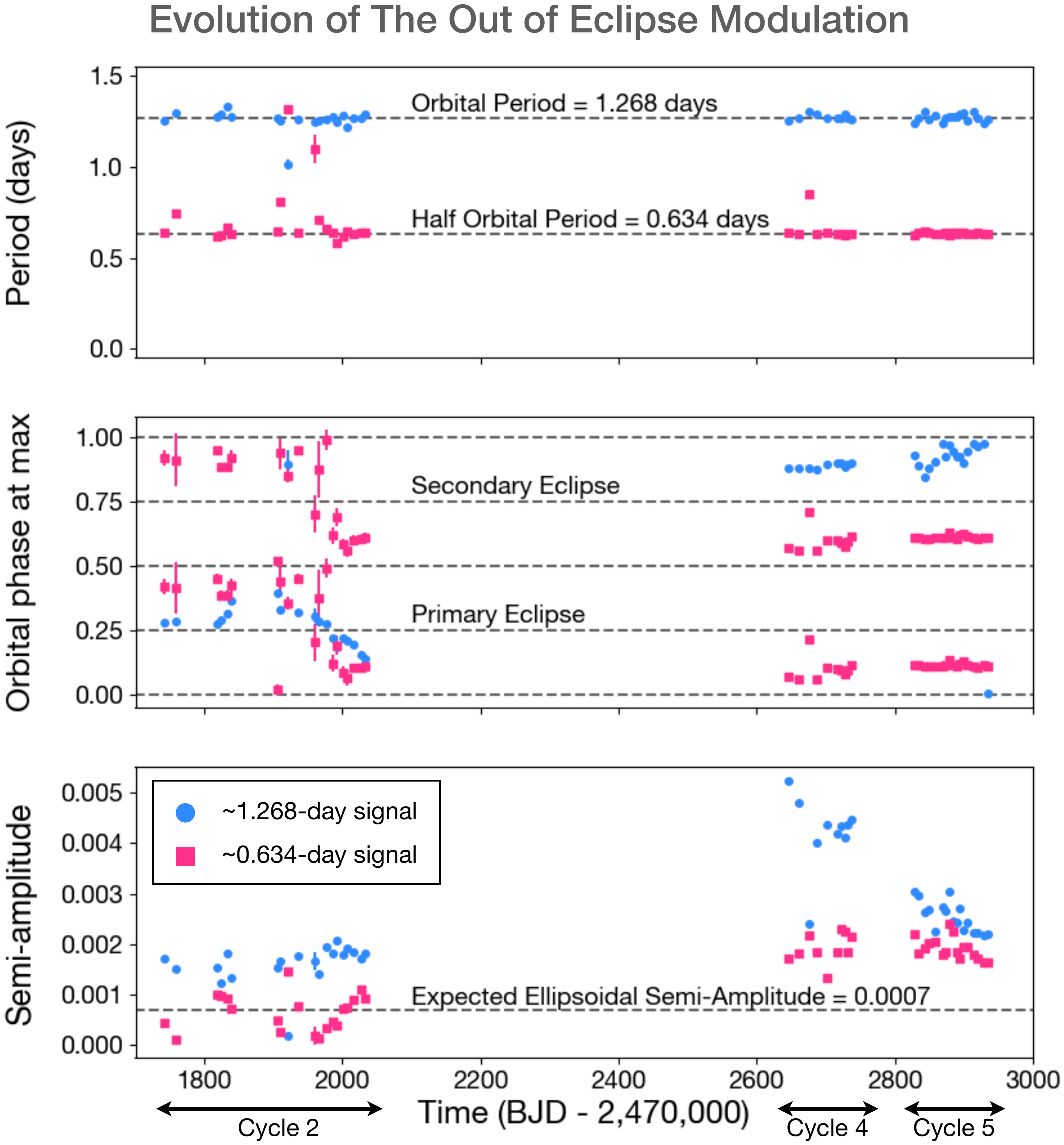}
    \caption{Activity variation over time. Each data point comes from the two sine curve fit to a 5.072-day segment ($=4\times P_{\rm bin}$) of the light curve, as demonstrated in Fig.~\ref{fig:fitted_sine_curves}.  Errorbars are not visible for most points. {\bf Top:} period of the fitted sinusoids. The strongest frequency is at $P_{\rm bin}$, shown as blue circles. There is a weaker but significant frequency at $1/2P_{\rm bin}$, shown as pink squares. {\bf Middle:} orbital phase of the binary where the sinusoid is at a maximum. At each time step the $1/2P_{\rm bin}$ frequency has two max phases within one CM Dra orbital period. {\bf Bottom:} semi-amplitude of the fitted sinusoids. For comparison, we show the expected ellipsoidal variation amplitude (Eq.~\ref{eq:ellipsodal_amplitude}), which matches in TESS cycle 2 but not in cycles 4 and 5.}
    \label{fig:spot_modulation}
\end{figure}

In Fig.~\ref{fig:spot_modulation} we show the period (top), phase (middle) and amplitude (bottom) of the two sinusoids fitted to each $4\times P_{\rm bin}$ segment. The periods are consistently close to either $P_{\rm bin}$ (1.268 days, blue circles) or $1/2P_{\rm bin}$ (0.634 days, pink squares). The phases and amplitudes of both the $P_{\rm bin}$ and $1/2P_{\rm bin}$ signals are seen to change, both between cycles and within cycles. We analyse what we believe to be the source of these two signals in Sect.~\ref{subsec:modulation_cause}.

\section{ANALYSIS \& DISCUSSION}\label{sec:analysis}

\subsection{What Causes The Light Curve Modulation?}\label{subsec:modulation_cause}

We attribute the 1.268 day signal (Fig.~\ref{fig:spot_modulation}) to star spots, as expected from two tidally locked M dwarfs with a 1.268-day orbital period. We do not know which star is producing the spot signal, but it is likely that both stars have a high spot coverage and the 1.268 signal is an amalgamation of the two given that the two stars are so similar in size. The spot modulation changes both in phase and amplitude, which we analyse in Sect~\ref{subsec:activity_cycle} in the context of a possible activity cycle. The trickier question is  what causes the $1/2P_{\rm bin}$-day frequency?

For a tight binary we would typically ascribe a frequency at $1/2P_{\rm bin}$ to ellipsoidal variation. This is the  tidal deformation of the star such that at different orbital phases we see different surface areas, and hence different brightnesses. The ellipsoidal signal should have three attributes: 1) a period at exactly $1/2 P_{\rm bin}$, 2) a phase such that the signal is maximised at orbital phases 0.0 and 0.5 (i.e. it is minimised during eclipses at phases 0.25 and 0.75) and 3) a constant amplitude:

\begin{equation}
\label{eq:ellipsodal_amplitude}
A_{\rm ellip} \approx 2\times\alpha_{\rm ellip}\frac{M_{\rm B}}{M_{\rm A}}\left(\frac{R_{\rm A}}{a_{\rm bin}}\right)^3,
\end{equation}
where
\begin{equation}
\alpha_{\rm ellip} = 0.15\frac{(15 + u)(1+g)}{3-u}
\end{equation}
is calculated using the stellar gravity darkening coefficient $g$ and the limb darkening coefficient $u$ within the TESS bandpass. Equation~\ref{eq:ellipsodal_amplitude} is taken from \citet{mazeh10}. We added a factor of 2 to account for the fact that we have two identical stars inducing ellipsoidal variation in each other, as opposed to the star-exoplanet binaries considered in \citet{mazeh10}. We calculate $A_{\rm ellip}\approx0.0007$.

With this in mind, we consider the following possibilities for this $1/2 P_{\rm bin}$-day signal:

\begin{enumerate}
    \item {\bf Ellipsoidal variation:} Within Cycle 2 we see the 0.634-day signal looks roughly as expected, with respect to its period, phase and constant amplitude. In Cycles 4 and 5, however, things look peculiar. The period remains clearly 0.634 days, yet the average amplitude has roughly doubled. Furthermore, the phase has changed to an average of roughly 0.6. This may seem close to the expected 0.5, but as demonstrated in Fig.~\ref{fig:fitted_sine_curves} the minimum is noticeably offset from the eclipses. We conclude that the power of the $1/2P_{\rm bin}$ frequency is not (at least predominantly) coming from ellipsoidal variation.
    \item {\bf Reflection or Doppler beaming:} These other binary-specific effects occur with a period of $P_{\rm bin}$, not $1/2P_{\rm bin}$, so they do not work.
    \item {\bf Dilution from a neighbouring star:} One might suggest that a change in amplitude  of the  variation could be caused by varying dilution from a neighbouring star which we have not accounted for. However, we can rule this out based on CM Dra being fairly bright (Jmag = 8.5) and any such time-dependent dilution would also affect the eclipse depths, which is not seen. Dilution could also not explain a change in phase. %Furthermore, all six sources within 1 arcmin of CM Dra are too dim (Vmag $> 11$) to induce such significant fluctuations in the light curve.
    \item {\bf Each star has a different rotation rate:} If both stars are spotted but have different rotation rates then we would expect multiple frequencies. However, we are unaware of any tidal mechanisms in a tight twin binary that would produce a $1:1$ spin-orbit ratio in one star and a $2:1$ spin-orbit ratio in the other.
    \item {\bf Differential rotation:} Spots can manifest as multiple frequencies if they occur at different latitudes in a star with differential rotation. This is seen on the Sun. However, to have one latitude of CM Dra rotate at $P_{\rm bin}$ and another at precisely $1/2P_{\rm bin}$ seems implausible. 
    \item {\bf Spot clusters separated by 180$^{\circ}$ in longitude on a single star:} If spots are not uniformly distributed in longitude but rather in two distinct clumps on opposite sides of the star then this would manifest as a $1/2P_{\rm bin}$ signal.
    \item {\bf Spot clusters at sub-stellar points:} If both stars have an overabundance of spots near the substellar point then we would predict two things. First, this would produce a $1/2 P_{\rm bin}$ periodicity since the observer would see each cluster once during an orbital period, separated by half an orbital period. Second, we would expect the $1/2 P_{\rm bin}$ signal to have a maximum flux near orbital phases 0.0 and 0.5 (i.e. a minimum flux near eclipse at phases 0.25 and 0.75). This is close to what we see in Fig.~\ref{fig:spot_modulation}. 
\end{enumerate}

\begin{figure}
    \centering
    \includegraphics[width=0.40\textwidth]{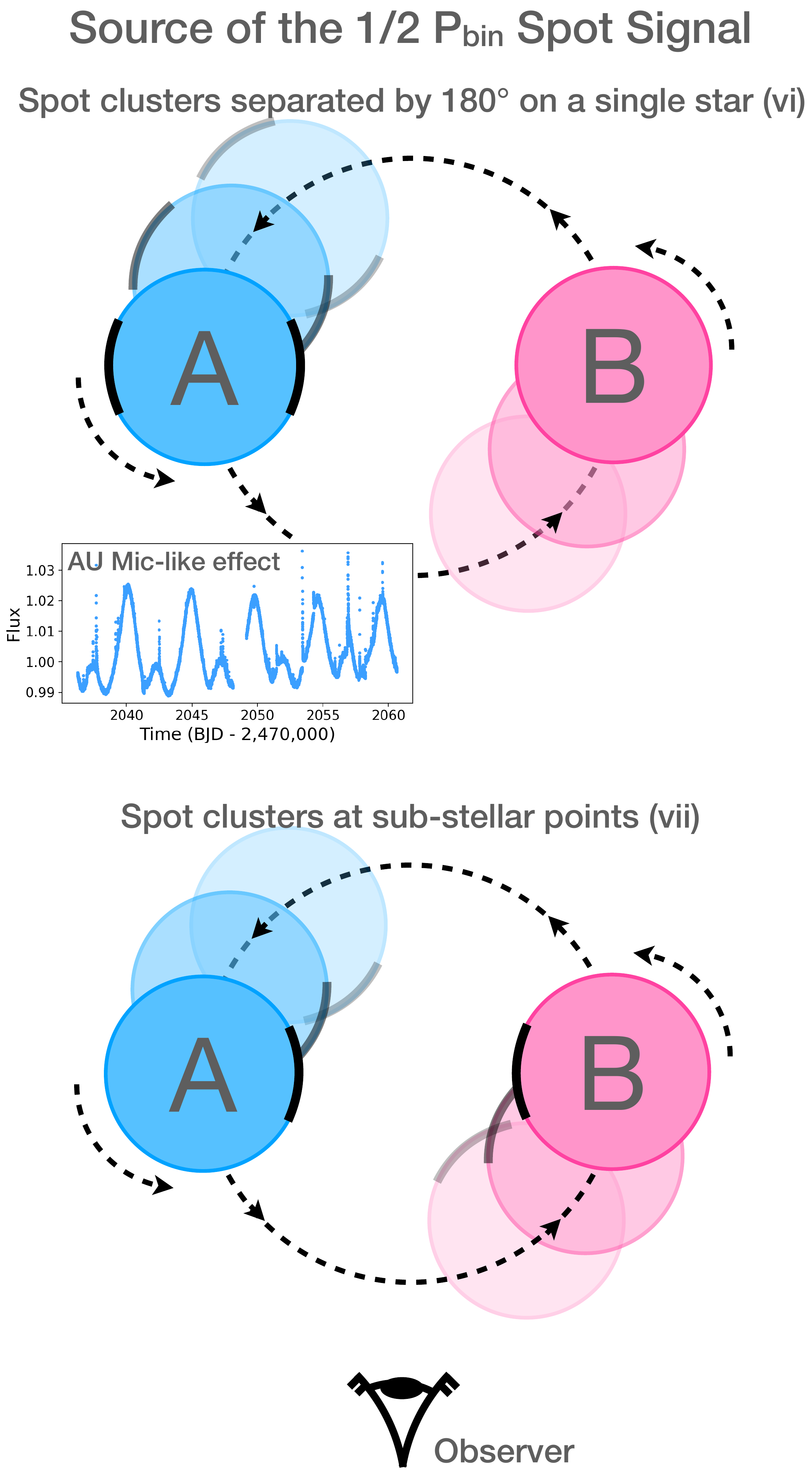}
    \caption{The two most plausible explanations for a spot modulation signal at $1/2P_{\rm bin}$. These are explanations (vi) and (vii) from our list of seven in Sect.~\ref{subsec:modulation_cause}. The primary (A, blue) and secondary (B, pink) stars are assumed to have spots across many longitudes, which will produce the most prominent signal at $P_{\rm bin}$. Additionally, there is an over-abundance of spots clustered in longitude, denoted in this diagram by a black strip. {\bf Top:} one star (we do not know which) has concentrations of spots separated in longitude by $180^{\circ}$. This effect was seen prominently in the single star AU Mic, shown in the inset. {\bf Bottom:} both stars have spot clusters near the sub-stellar point. The sub-stellar point is constant due to tidal locking.} 
    \label{fig:half_binary_spot_signal}
\end{figure}

The two most plausible explanations are (vi) and (vii), which we illustrate in Fig.~\ref{fig:half_binary_spot_signal}. Explanation (vi), with spot clusters separated by $180^{\circ}$ on a single star, is plausible because this effect has been seen on single stars such as  AU Mic, which is another BY Draconis variable (Fig.~\ref{fig:half_binary_spot_signal} inset). There is clearly a primary periodicity at 4.8 days, with a secondary periodicity at half that. \citet{Martioli2021,Szabo2021} also attributed this to $180^{\circ}$-separated spots. \citet{Skarka2022} noted this effect in 31 magnetic rotator (ROTM) variables out of a few hundred A-F spectral type stars. It was also noted that this effect may be confused with ellipsoidal variables of non-eclipsing binaries. With TESS we also see this $1/2$ periodicity effect in BY Draconis itself, which we will study in a future paper. 

Explanation (vii), which is only for binary stars, is also plausible. It was  proposed by \citet{Simon1980,vandenOord1988,Gunn1997} that there may be magnetic reconnection field lines that connect two stars in a tight, tidally locked binary, and this may increase spot coverage near the sub-stellar points, and hence a reduced flux near eclipse. We note that this effect would be easily confused with ellipsoidal variation. The main difference is that whilst ellipsoidal variation must have a minimum exactly during eclipse, it is reasonable that the spot cluster may not be precisely on the sub-stellar point, and hence the minimum can be just near eclipse.

The fact that CM Dra is a binary does indeed confuse things. In the future if we see the phase of the $1/2P_{\rm bin}$ signal change significantly then this would favour explanation (vi), since this explanation has no restriction on the spot phase whereas (vii) dictates the phase with respect to the eclipses. We ultimately leave it to a future study to construct a physical evolving spot model that is a good fit to the CM Dra data.

\subsection{Flare Rate}\label{subsec:flare_rate}

Our time-series spans  1198 days. By cutting out gaps between and within sectors, there are  328 days worth of observations. With a total of 163 flares, we deduce a simple flare rate of 0.5 flares per day. We recall that \citet{Lacy1977} calculated a flare rate between 0.48 and 1.2 flares per day based on a single flare, which is compatible with our result from  163 flares. 

One important subtlety is that this is for the CM Dra binary as a whole. If we are to compare this with flare rates of single M dwarfs we must account for the fact that it is a binary and hence two potential sources of flares. We will argue in Sect.~\ref{subsubsec:orbital_phase} that the TESS data demonstrate flares occur on both stars. This is not surprising given that the stars are practically twins. We therefore approximate the flare rate on each individual M dwarf as 0.25 flares per day, i.e. half the total rate.

\begin{figure}
    \centering
    \includegraphics[width=0.50\textwidth]{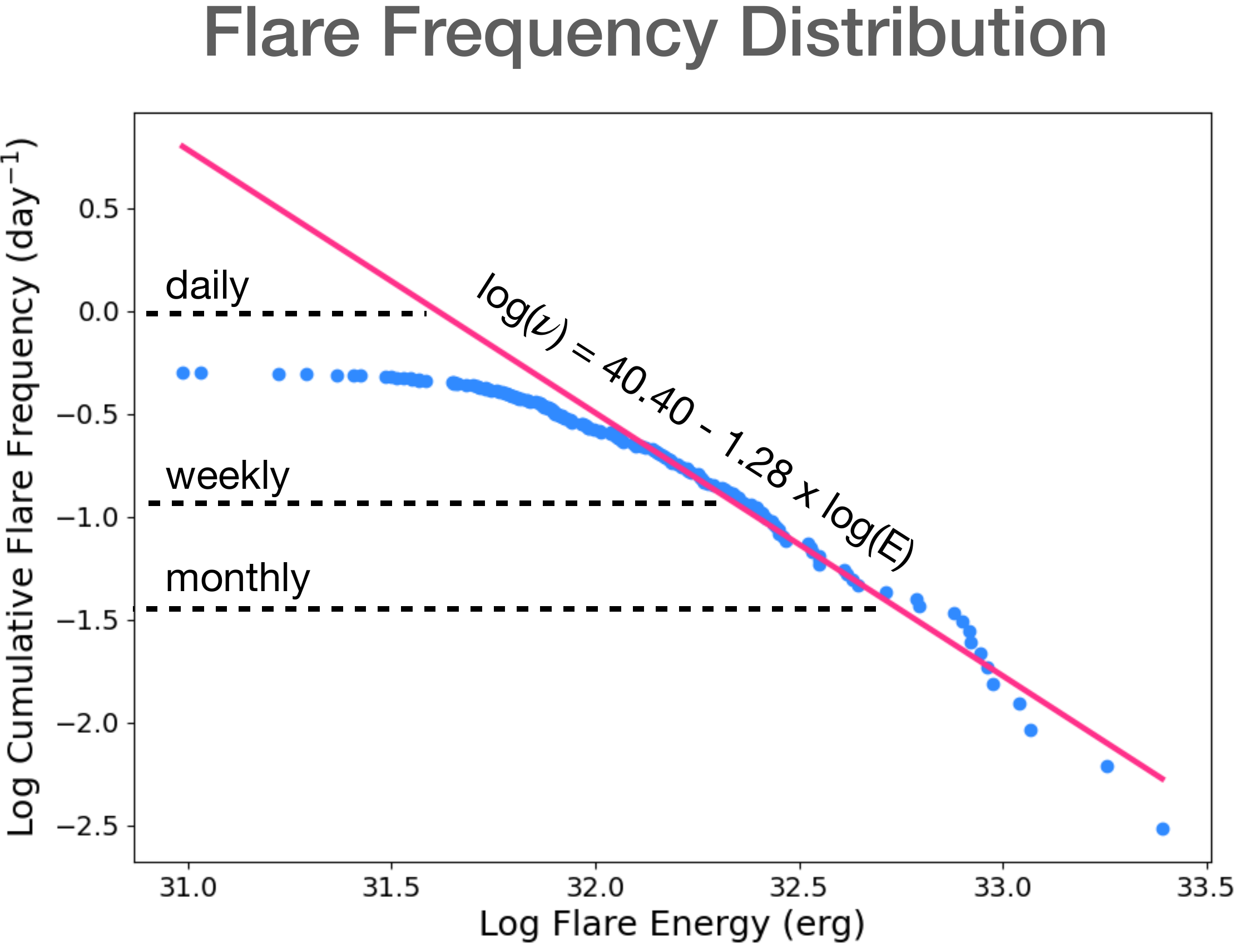}
    \caption{Flare frequency distribution, with both horizontal and vertical axes in log scales. The vertical axis is a cumulative rate of flares up to a certain energy. That is to say, flares with an energy of $10^{31.7}$ erg and above occur on a daily basis, flares over $10^{32.3}$ erg occur on a weekly basis, flares over $10^{32.7}$ occur on a monthly basis, and so on.}
    \label{fig:flare_frequency_distribution}
\end{figure}

A more sophisticated way of quantifying the flare rate is to calculate the flare frequency distribution (originally proposed by \citealt{Lacy1976}):

\begin{equation}
\log_{10}(\nu) = \alpha+\beta\log_{10}(E),
\end{equation}
where $\nu$ is the flare frequency and $E$ is the flare energy. For our fit $\alpha=40.40$ and $\beta=-1.28$. This is plotted in Fig.~\ref{fig:flare_frequency_distribution}. This accounts for  flares being produced with a distribution of energies. It also accounts for detection limitations, i.e. our flare sample will not be complete at low flare energies. To assist interpretation of Fig.~\ref{fig:flare_frequency_distribution} the horizontal lines demarcate the energy of flares that occur on a daily, weekly and monthly basis. The pink line is a power law fit. It is fitted to the higher energy flares (in this case $>10^{32}$ erg) because they are the easiest to detect, and hence we believe the detections are complete in this parameter space. At lower energies the curvature of the flare frequency distribution indicates that the flare sample is not complete because of limits in detection sensitivity when it comes to low energy flares.

The flare frequency distribution allows for comparisons with other stars. \citet{MacGregor2021} studied our nearest stellar neighbour, Proxima Centauri, and derived a flare frequency distribution of $\log_{10}(\nu) = 27.2 - 0.87\log_{10}(E)$. Compared with CM Dra, Proxima Cen has a smaller $\alpha$ but a larger $\beta$, i.e. a less steep negative slope. This means that CM Dra has more frequent low-energy flares, but Proxima Cen has more frequent high-energy flares.

%\citet{Stelzer2022} discovered 18 flares in CM Dra using only the first available TESS sector (16).

\subsection{Flares as a Function of Phase}\label{subsec:flare_phases}

\begin{figure}
    \centering
    \includegraphics[width=0.50\textwidth]{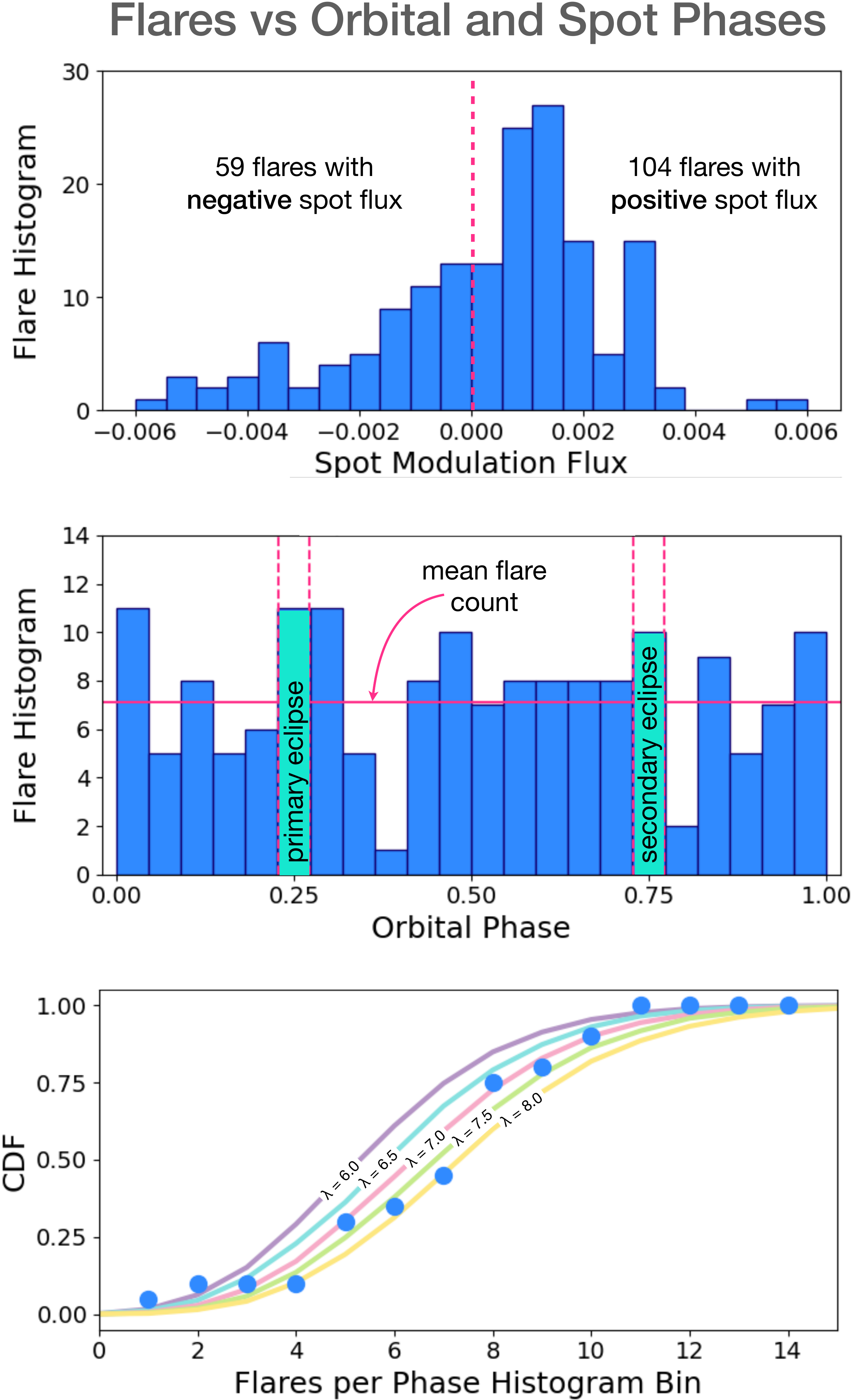}
    \caption{{\bf Top:} flares as a function of the spot modulation flux at the time of flare. The spot modulation flux is taken from the \textsc{Wotan} fit after removing parabolic TESS orbit trends. One might expect more flares during  negative flux, because that means more spots are facing the observer. Instead, the opposite is seen. {\bf Middle:} flares as a function of orbital phase. One might expect less flares during the primary and secondary eclipses, but again, the opposite is seen. {\bf Bottom:} cumulative distribution function (CDF) of flares within each orbital phase histogram bin from the middle plot. Flares within the eclipses are excluded. For comparison, we show different poisson distributions, parameterised by $\lambda$.}
    \label{fig:flare_histograms}
\end{figure}

We are interested in the flare distribution with respect to two phases: the orbital phase and the spot modulation phase. Even though the orbital and main spot periods are the same, the spot phase changes over the TESS timeseries. We therefore analyse the two phases separately.

\subsubsection{Spot Phase}\label{subsubsec:spot_phase}

Since flares and spots are expected to be associated, one might  expect that more flares will occur when the spot modulation flux is negative. This is because a negative flux implies a higher-than-average concentration of spots facing the observer, and hence associated flares would also face the observer. However, we see in Fig.~\ref{fig:flare_histograms} (top) that in fact the opposite is true:  104 flares (64\%) coincide with a positive spot modulation flux. The spot modulation flux is taken as the \textsc{wotan} trend fitted in Sect.~\ref{sec:data_acquisition} with the parabolic TESS-orbit trends removed according to Fig.~\ref{fig:parabolas}. There is a slight asymmetry in the spot modulation flux overall - 52\% of all data points are positive - but this is not enough to account for the asymmetry in Fig.~\ref{fig:flare_histograms} (top). 

We quantify the significance of our apparent correlation between flare occurrence and positive spot flux using a simple bootstrap test. We randomly draw  163 flares across our time series and test if  104 or more flares correspond to a positive flux. We repeat this process $10^4$ times. From this derive a p-value of  0.0083.  This implies good evidence of a positive correlation between the flare rate and positive spot modulation.

 As a second test, we conduct a one-sample Student t-test, comparing the distribution of the spot flux of the  163 flares to a normal distribution with a mean of zero. We return $t_{\rm stat}=1.942$  and a p-value of 0.054. We interpret this as moderate evidence that the flares are not normally distributed around zero spot modulation flux. We also note a caveat that an assumption of the Student t-test is that events are independent, whereas flares may exhibit ``sympathetic flaring'', where the production of one flare may increase the likelihood of subsequent flares shortly after \citep{Moon2002,feinstein2022}. We do not analyse sympathetic flaring in CM Dra.

%An earlier version of this bootstrap test (the first version of this paper published on arXiv) was missing the final 3 sectors of Cycle 5 and derived a p-value of 0.0315. This would be considered much weaker evidence of any correlation.

 Overall, our data provide preliminary evidence that on CM Dra flare rates are {\it inversely} proportional to spot occurrence, which is contrary to the expectation. However, we recommend repeating this analysis with future TESS data before drawing strong conclusions. Fortunately, TESS will observe CM Dra for 7 sectors in Cycle 6, starting March 2024. As previously stated, one might expect more flares when there is a {\it negative} spot modulation  (i.e. negative flux means more spots are visible and flares are expected to be more likely near spots). If there truly are more flares during {\it positive} spot modulation, then this would have interesting implications for the models of stellar activity and surface inhomogeneities. Some recent literature studies. \citet{Dal2011,Hawley2014,Doyle2018} found found no correlation between flare rate and rotational phase. \citet{Roettenbacher2018} found no correlation for the stronger flares and only weak correlation for the weaker flares. 
 
 One possible explanation is that the flares are indeed correlated with dark starspots, but they are also correlated with bright spots known as faculae/plages. Starspots are dark spots on the stellar photosphere, resulting in a flux deficit. Faculae are conversely bright spots on the photosphere. Plages are corresponding bright spots observed higher up in the stellar chromosphere. Our Sun, for example, actually gets brighter when there are more sunspots. Whilst this might seem unintuitive, this is because the presence of spots is accompanied by the presence of faculae/plages. For the Sun, the excess brightness of plages/faculae outweighs the flux defecit of the spots, resulting in a net increased brightness of 50\% during peak stellar activity in the Solar cycle \citep{Chapman1997}. It would be of interest to test this trend in other M-dwarf stars, but we leave that for future studies.

\subsubsection{Orbital Phase}\label{subsubsec:orbital_phase}

\begin{figure*}
    \centering
    \includegraphics[width=0.99\textwidth]{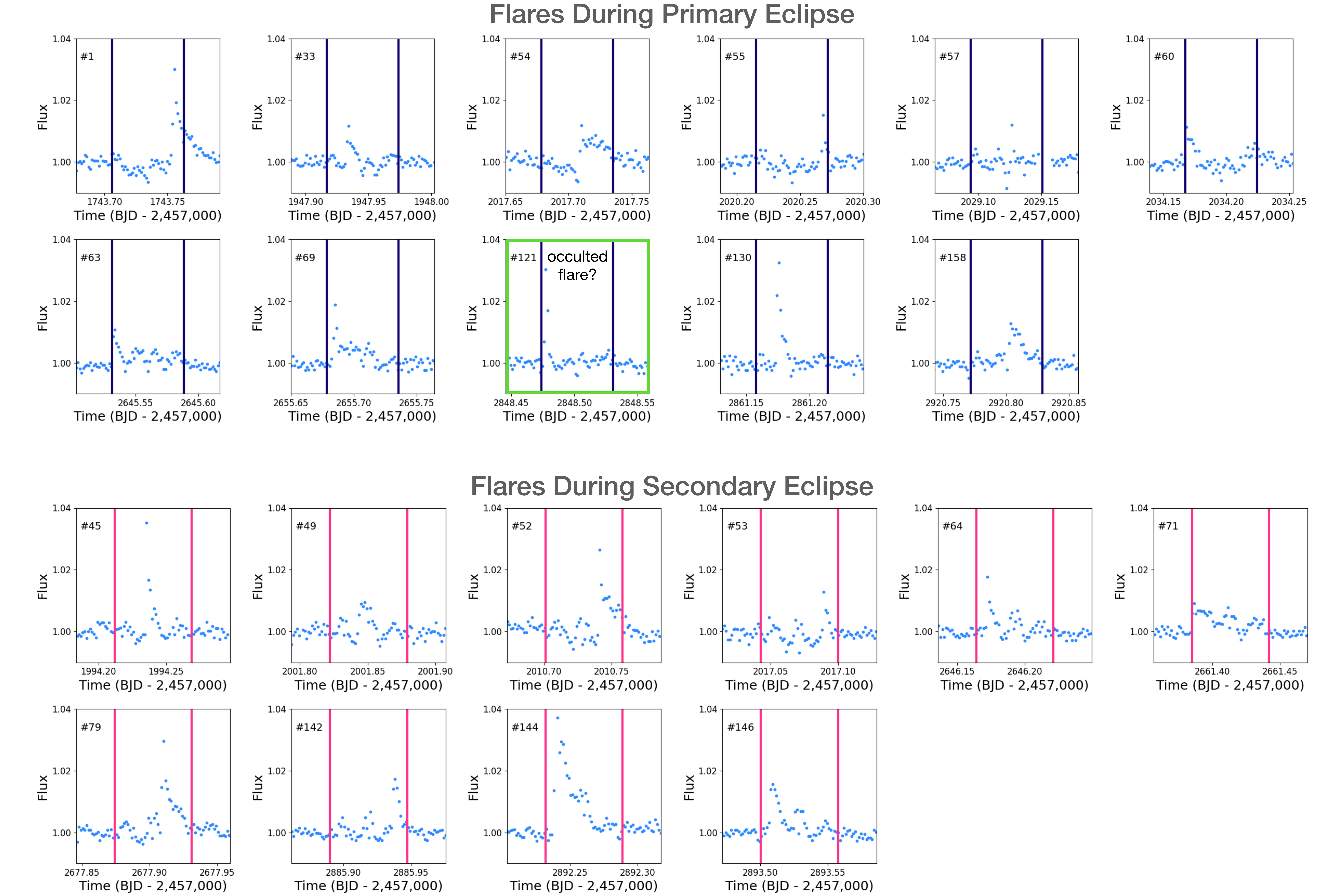}
    \caption{Flares where the peak of the flare (as detected by \textsc{stella}) occurs within a primary (top) or secondary (bottom) eclipse. The vertical lines demarcate the eclipse duration. The numbers in the top left corners indicate the flare number, out of the 125 flares discovered.}
    \label{fig:in_eclipse_flares}
\end{figure*}

In Fig.~\ref{fig:flare_histograms} (middle) we plot a histogram of the flare count as a function of orbital phase. The choice of bins is such that the bins centered at phase 0.25 and 0.75 (highlighted in aqua) correspond to an eclipse duration (1.37 hours). The mean flare count across all non-eclipse bins is 7.10. By eye, there is no obvious phase preferred by flares. In Fig.~\ref{fig:flare_histograms} (bottom) we calculate the cumulative distribution function (CDF) across all non-eclipse histogram bins. We compare this with a Poisson distribution for five different $\lambda$ parameters. We see that the non-eclipse flares can be modelled roughly by a Poisson distribution with $\lambda=7.5$. We therefore conclude that the flare distribution as a function of orbital phase is flat aside for Poisson noise, i.e. there is no favoured orbital phase. A flat orbital phase distribution of flares is in contrast to the spot distribution, which seemingly has some clustering (Sect.~\ref{sec:spots}). 

%It was also proposed by \citet{Simon1980,vandenOord1988,Gunn1997} that there may be magnetic reconnection field lines that connect two stars in a tight, tidally locked binary, and this may increase flare rates near the sub-stellar points. In this case there would be an increase of flares near phases 0.27 and 0.75, but we do not see that.

We treat the flare counts within the primary and secondary eclipse separately because they should not be representative of the rest of the orbit. Within each eclipse one of the stars is almost completely obscured. Depending on the distribution of flares (longitude, latitude and across both stars) we should expect some reduction during eclipse.  {\it We do not see a reduced flare rate during eclipse}.

Seven flares are seen during primary eclipses and nine are seen during secondary eclipses.  These counts are  higher than the mean of 7.10 across non-eclipse phases. If we take the 22 bins and fold them across a phase of 0.5, there are  21 flares in eclipse, whereas there out of eclipse bins have $14.2\pm 3.5$ flares, where $\sigma=3.5$ is the standard deviation of the bin counts. We discuss several possible flare distributions and their compatibility with our surprisingly high in-eclipse flare rate:

\begin{enumerate}
    \item {\bf Uniform flares across both stars.} If flares are spread across all latitudes and longitudes then we should expect a 50\% reduction in the flare rate at eclipse totality, since the background star is completely blocked and the only flares can come from the nearside of the foreground star. If we marginalise over the entire V-shaped eclipse (Fig.~\ref{fig:light curve}, right), accounting for ingress and egress, then there should be a 25\% reduction in the flare rate within the entire eclipse bin. Therefore, we should expect $10.65$ flares across both eclipses. Our measurement of 21 flares is incompatible with this by $approx 3\sigma$.
    \item {\bf Flares concentrated at sub-stellar points.} \citet{Simon1980,vandenOord1988,Gunn1997} proposed that there may be magnetic reconnection field lines that connect two stars in a tight, tidally locked binary, and this may increase flare rates near the sub-stellar points. This might be connected to the possible spot clustering near the sub-stellar points, illustrated in Fig.~\ref{fig:half_binary_spot_signal}. One would therefore expect more flares at phases 0.25 and 0.75, hence explaining our high in-eclipse flare count even if some of them are obscured. However, a concentration of flares near sub-stellar points should not produce the flat distribution of flares seen outside eclipse, so this scenario seems to be unlikely.
    \item {\bf Only one star flares.} In this case one eclipse would have no flare reduction and the other would have a 50\% reduction (marginalised over the entire eclipse bin). Since there is a similar number of flares during each eclipse (seven and nine) and both are above the mean out of eclipse flare count, this scenario also seems unlikely.
    \item {\bf Confusion with spot crossings.} If a spot is occulted during an eclipse then there is an upwards bump in the light curve since a relatively dark region is being blocked. The morphology would be different to a typical flare, which has a sharp rise and exponential decay, but we speculate that \textsc{stella} could possibly confuse the events. This would imply that some of our 16 in-eclipse ``flares'' are in fact spot crossings, and hence the true flare rate may actually be reduced. We do not favour this scenario because most of the events in Fig.~\ref{fig:in_eclipse_flares} truly look like flares.
    \item {\bf Polar flares.} Since CM Dra has such a small impact parameter ($b=0.11$), if the flares are predominately at high latitudes, and the two stars are spin-orbit aligned\footnote{Whilst spin-orbit alignment due to tides may be expected in CM Dra since it is such a tight binary (e.g. like EBLM J0608-59, \citealt{Kunovac2020}), to our knowledge it has not been measured. This may come in the future through the Rossiter-McLaughlin effect, although blended M dwarf spectra are difficult to precisely measure. Alternatively, we have seen that CM Dra is very spotted. The orbital obliquity might be measurable through spot-crossing events, although we have not noticed any definitive examples of these events. Spot crossings could also possibly be confused with in-eclipse flares.}, then there will be less blocking of the flares during eclipse. Therefore, in this scenario the in-eclipse flare count should be the same as the out of eclipse flare count. This explanation is the most compatible with our data, but we do not consider this solution conclusive.  A polar distribution of flares would affect the timing of flares observed during eclipse, i.e. we would see more flares at the start and end of the eclipse compared with in the middle. However, conducting a significant analysis on our small sample of 21 flares observed during eclipse is not possible, particularly when you consider the fact that both stars could flare and hence a flare in the middle of an eclipse could be on the nearside of the foreground, occulting star.
\end{enumerate}

Evidence for the latitudinal distribution of flares on M dwarfs is currently minimal. \citet{Ilin2021} discovered six flares on four M dwarfs with latitudes above $50^{\circ}$. This discovery came using rapidly-rotating stars ($P_{\rm rot}\approx$ hours) with energetic flares on a longer timescale. The rotation brings the flare in and out of view, modulating the light curve and allowing the latitude to be isolated. Our discovered flare distribution is consistent with polar flares, and hence the results of \citet{Ilin2021}. However, our evidence is only marginal and is a less direct method of constraining flare latitudes.

\citet{Huang2022} studied 12 M dwarf eclipsing binaries and found that only some showed a relationship between flare rate and orbital dependence. There were also a couple of targets for which the flare rate seemed to drop during eclipse. These results were not interpreted within the context of flare latitudes.

%Our 

%For M dwarfs there is preliminary evidence that spots \citep{Barnes2017} and flares \citep{Ilin2021} might be polar. This would likely be good for habitability, since the flux from the flares would be attenuated and the charged particles from flares would be directed away from the planet. Note that this assumes spin-orbit alignment for the star and planet, which works for the solar system and most exoplanets.

\subsection{Occulted Flares}\label{subsec:occulted_flares}

\begin{figure}
    \centering
    \includegraphics[width=0.45\textwidth]{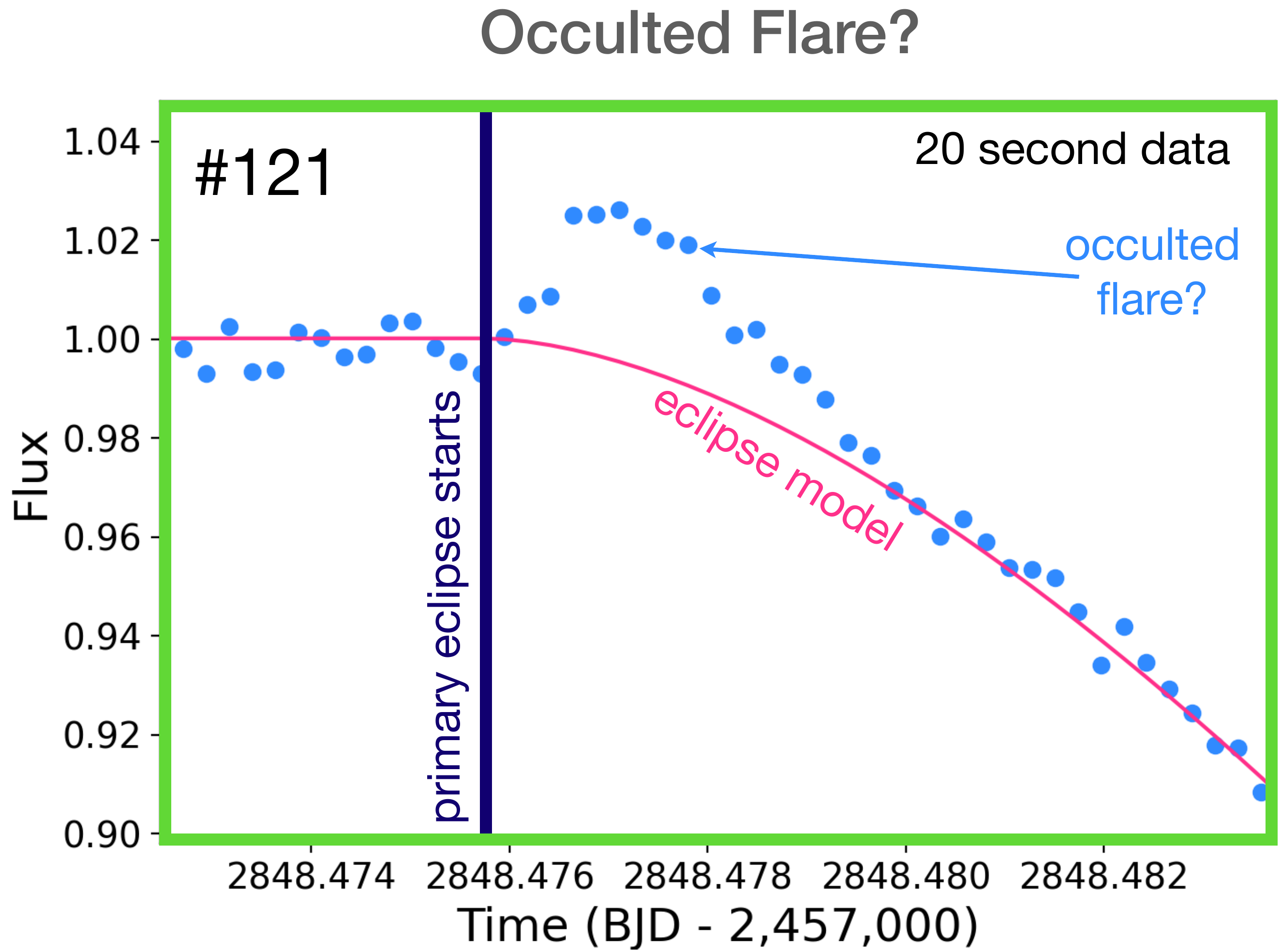}
    \caption{A candidate occulted flare as seen with 20 second cadence TESS data. This is flare \#121 out of 163 detected. It occurs at the start of a primary eclipse and has a very short duration ($\approx 200$ seconds), potentially indicative of it being occulted by the foreground secondary star. This flare is highlighted in green in Fig.~\ref{fig:in_eclipse_flares}.}
    \label{fig:possible_occulted_flare}
\end{figure}

\begin{figure}
    \centering
    \includegraphics[width=0.50\textwidth]{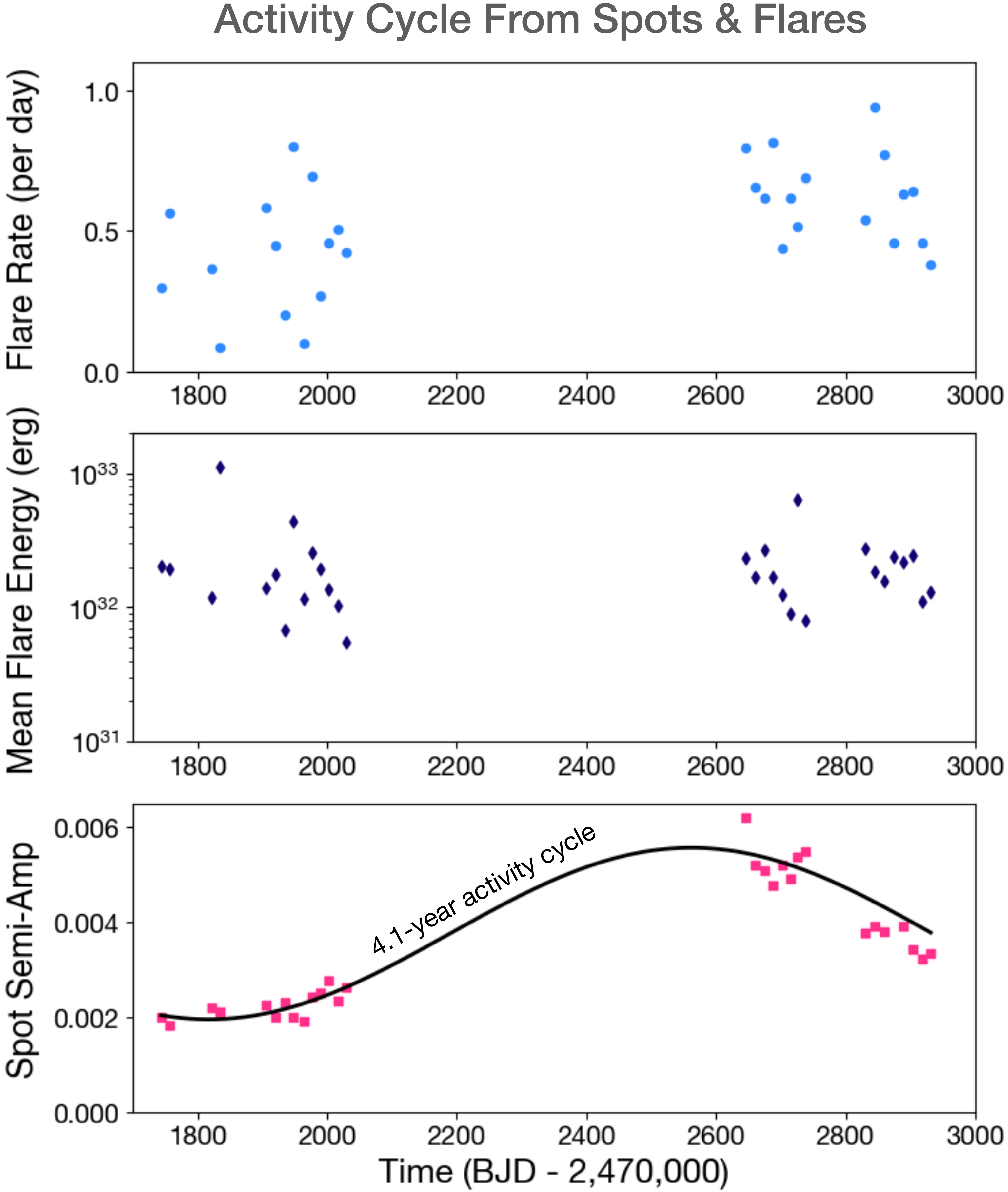}
    \caption{Activity cycle of CM Dra. All of the data points are calculated within a given 13.7-day TESS orbit. {\bf Top:} average flare rate. {\bf Middle:} average flare energy (erg). {\bf Bottom:} semi-amplitude of the out of eclipse spot modulation, calculated as half of the maximum peak to peak difference of the \textsc{wotan} trend within a given TESS orbit. This trend encompasses both the 1.268 and 0.634-day frequencies, but parabolic intra-orbit trends (Fig.~\ref{fig:parabolas}) have been removed. We see a significant increase in the amplitude of spot modulation over time, which is accompanied by a modest increase in the flare rate. The mean flare energy, however, does not vary. Based on the spot modulation we estimate a  4.1-year activity cycle.}
    \label{fig:activity_cycle}
\end{figure}

In Sect.~\ref{subsubsec:orbital_phase} we counted  21 flares during primary and secondary eclipses. Depending on the timing of the flare, it is possible that the flare will be occulted by the foreground star, imprinting a distinct signature on the light curve. We will investigate this effect in a dedicated future paper (Armitage et al., under review), but for now we conduct a brief investigation into CM Dra.

From Sect.~\ref{sec:flares}, the median surface area of the flares (at max flux) is $6.9\times10^{13}$ m$^2$. Naively assuming circular flare regions, this corresponds to a diameter of $4.7\times10^6$ m, which is $1.3\%$ of CM Dra A's diameter. During eclipse, the relative motion of the two stars is $\approx 150$ km/s, meaning that the entire flare region will be passed over a timespan of $\approx 30$ seconds. With 120-second cadence data the flare ingress/egress of the occultation would therefore be effectively instantaneous.

In Armitage et al.  (under review) we detail the phenomenon of flare occultations,  including the possible light curve morphologies. In the simplest cases though, if a flare is on the decay phase and is then occulted you will see a sharp and noticeable drop in flux, assuming the occultation occurs when the flare is still sufficiently bright. The main alternative is that the flare starts when it is being occulted by the foreground star, and subsequently the flare is revealed as the star passes by. In such a scenario the rapid egress of the occultation may be indistinguishable from the rapid rise of a standard flare, and hence we may not be able to deduce that the flare was occulted. 

In Fig.~\ref{fig:in_eclipse_flares} we plot all  21 flares discovered during eclipse. Most of the flares occuring within the eclipse do not show signs of being occulted. For this to be the case they must either occur on the nearside of the foreground star, or on nearside of the background star but on an uncovered region of its surface. Flare \#121, on a primary eclipse, may be occulted because there appears to be a sharp drop in flux, as opposed to the typical gradual drop in flux (most clearly evidenced in flares \#1 and \#79). We highlight this flare in green and then plot the 20 second data separately in Fig.~\ref{fig:possible_occulted_flare}. For flare \#121 we also plot the 20 second data in  (highlighted in green), where in this plot the eclipse model is shown but not subtracted. There appears to be a typical sharp rise in the flare and then a decay over $\approx 200$ seconds. This is a little longer than the $\approx30$ second calculation from before, but we emphasise that that was a rough estimate. If we compare flare \#121 to  say flare \#1 we see that both stars have a similar peak amplitude, yet flare \#1 takes a much longer $\approx3500$ seconds to decay.

One challenge with CM Dra though is that since effectively all latitudes of each star are blocked at some point of the eclipse, the observation of an occulted flare does not necessarily identify its latitude. However, since the potential occultation of flare \#122 occurs early in the eclipse, the flare must have been near the star's equator (assuming spin-orbit alignment). This is in contrast with our deduction of polar flares in Sect.~\ref{subsubsec:orbital_phase}, but ultimately neither piece of evidence is conclusive.

An occulted flare from a transiting exoplanet more clearly identifies the flare latitude because planets only cover a narrow transit chord. Although, one does still have the question of the spin-orbit alignment of the system. Flares may also have complex shapes  irrespective of any occultations (e.g. if there are multiple small flares simultaneously). We leave a thorough analysis of occulted flares to Armitage et al. (in prep).

\subsection{Activity Cycle}\label{subsec:activity_cycle}

On the Sun the occurrence of flares and spots varies on an 11-year cycle. We test this in CM Dra in Fig.~\ref{fig:activity_cycle} by plotting the flare rate, mean flare energy and spot modulation semi-amplitude over time. The data is split up to individual 13.7-day TESS orbits (two per sector).

The spot modulation has the strongest variation. This is essentially the same signal as shown in Fig.~\ref{fig:spot_modulation} (bottom) except the 1.268 and 0.634-day signals are combined. There is a slight increase within cycle 2. The jump from Cycle 2 to Cycle 4 corresponds to a large increase in flux, followed by a decrease throughout Cycle 4 and into Cycle 5. We can fit a  4.1-year activity cycle to these data, but we stress that a longer time-series is needed to properly constrain this. There is also ambiguity arising from the fact that we cannot assign spots to individual stars, and it is not guaranteed that CM Dra A and CM Dra B would have synchronised and identical activity cycles. The flare rate is seen to increase slightly over time, but the mean flare energy does not seen to change.

\subsection{Is CM Draconis Still Inflated?}\label{subsec:inflation}

\begin{figure}
    \centering
    \includegraphics[width=0.50\textwidth]{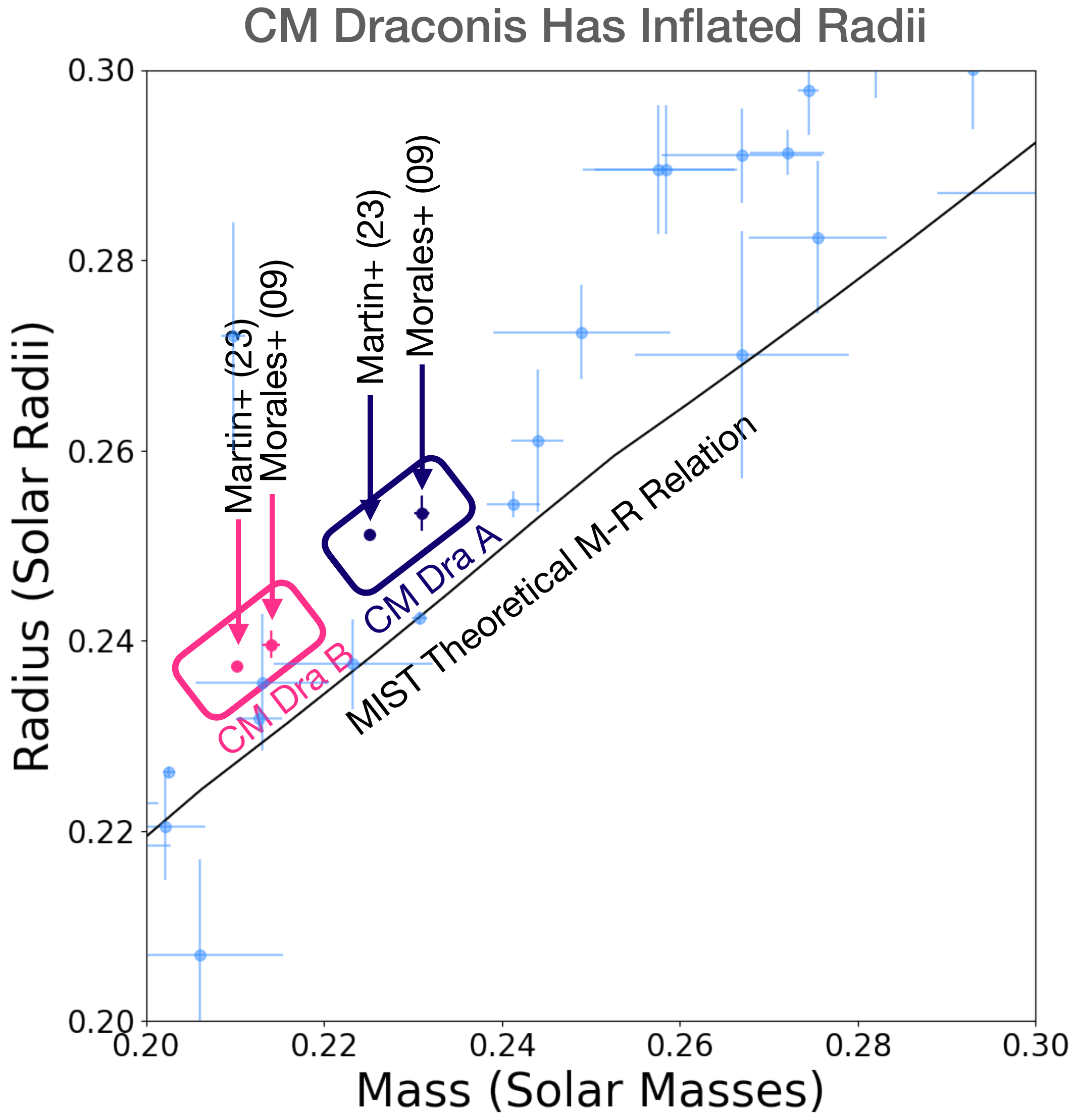}
    \caption{Mass-radius relationship between precisely-characterised M dwarfs in the literature (light blue), CM Dra A (navy) and CM Dra B (pink). We annotate our CM Dra values in comparison with the slightly larger \citet{morales2009} values. For our values the errorbars are not discernable at this scale. The difference between our values and \citet{morales2009} is discussed in Sect.~\ref{sec:eclipse_fitting}. The solid line is from the MESA Isochrone and Stellar Track (MIST) models \citet{Dotter2016}. The fact that CM Dra A and B have larger radii than expected by models is referred to as ``radius inflation''. All literature values have mass and radius fractional errors better than 5\%. Their values are taken from the review article \citet{Maxted2023}.}
    \label{fig:mass_radius}
\end{figure}

Yes, with our more precise parameters, the CM Dra radii are still inflated relative to stellar models\footnote{As a matter of semantics, one might argue that the radii are what they are and it is the models that are wrong, and hence it is really a ``model deflation'' problem.}. This is demonstrated in Fig.~\ref{fig:mass_radius}, where we plot the mass and radius of CM Dra A and B from both our work and \citet{morales2009}, along with some other precise M dwarf measurements from the literature \citep{Parsons2018,Duck2023}. The comparative model is from the Mesa Isochrone and Stellar Track (MIST, \citealt{Dotter2016}). Several well-characterised M dwarfs have larger (inflated) radii than expected at given masses. 

This MIST model is with a metallicity [Fe/H] $=0$. One may argue that deviations from the model are a result of a different metallicity. However, in the case of CM Dra this argument exacerbates the inflation; \citet{Terrien2012} derived [Fe/H] $=-0.30\pm0.12$ for CM Dra, and at a sub-solar metallicity we would expect an even {\it smaller} radius. \citet{Feiden2014inflation} suggest that the magnetic activity of CM Dra may be the cause of its inflation. Whilst we do not explicitly test this effect in this paper, this is further motivation to study CM Dra's stellar activity.

\subsection{Probing Activity Through Spectroscopy}\label{subsec:activity_spectroscopy}

We have solely used photometry to observe flares, but they can also be observed spectroscopically by detecting emission lines. \citet{Metcalfe1996} observed emission lines for Mg I and Fe II for CM Dra. Based on the radial velocity, the flares could be attributed to the primary star A, which is advantageous compared with photometry for which it is typically ambiguous which star caused any given flare, particularly for near twins like CM Dra. One may also use spectroscopy to calculate spot filling factors \citep{Somers2015,Cao2022}, where the spectrum is fitted with an additional spot component. Combined with a measurement of the projected rotational velocity $v\sin I$, this technique may reveal the spot latitude. Whilst we do not assess CM Dra's activity spectroscopically, we encourage future studies, in particular if they are simultaneous with future photometry from TESS or the PLAnetary Transits and Oscillations of stars (PLATO).

\section{CONCLUSION}\label{sec:conclusion}

We have studied the benchmark M dwarf eclipsing binary CM Draconis using  15 sectors of TESS data and archival radial velocities. The photometry is vastly improved in precision and baseline compared with the previous ground-based photometry. From this we derive radii at a precision of $\approx0.06\%$, which is an order of magnitude improvement on existing measurements, which were already amongst the best known for M dwarfs. The masses also increase in precision by a factor of $\approx 4$, now characterised to $\approx 0.12\%$. With updated parameters, CM Dra A and B are once again the most precisely characterised M dwarfs known. They also still have inflated radii relative to predictions from stellar models.

Our long baseline  (328 days spread over  1198 days) and short cadence (120 seconds) allows an unprecedented study of the flares and spots on CM Dra. We show strong out-of-eclipse modulations of the light curve with periodicities at $P_{\rm bin}$ and $1/2P_{\rm bin}$. We attribute these to spots in a tidally locked binary, where there is likely either a clustering of spots on opposite sides of one star, or both stars have spots concentrated near the sub-stellar points.

We discover  163 flares, at a rate of 0.5 flares per day, and find no preference for orbital phase. We argue that this may be indicative of polar flares, which would be in contrast to the equatorial flares seen on the Sun. Flares are also  more likely to occur when the photometric modulation of the light curve is positive. This implies either an anti-correlation between flares and dark sunspots, or a positive correlation between flares and bright plages/faculae.

We also discover  changes to the spot activity over a 4.1-year, which may be indicative of an activity cycle similar to the Sun. The flare rate is also seen to change slightly over this activity cycle, but no variation is seen in the average flare energy. Additional data from TESS,  including 7 sectors in the upcoming Cycle 6,     will improve our flare and spot statistics and provide a better probe of any activity cycles.

\section*{Acknowledgements}

 Support for this work was provided by NASA through the NASA Hubble Fellowship grant HF2-51464 awarded by the Space Telescope Science Institute, which is operated by the Association of Universities for Research in Astronomy, Inc., for NASA, under contract NAS5-26555. This research was carried out in part at the Jet Propulsion Laboratory, California Institute of Technology, under a contract with the National Aeronautics and Space Administration (80NM0018D0004)

\section*{Data availability}
We provide all light curves (raw, detrended and eclipse model-subtracted) online.

\bibliographystyle{mnras}
\bibliography{references}

\bsp	% typesetting comment
\label{lastpage}
\end{document}